# Field and stress-tunable microwave composite materials based on ferromagnetic wires

Dmitriy Makhnovskiy*) and Larissa Panina

*School of Computing, Communications and Electronics, University of Plymouth, Drake Circus, Plymouth, Devon PL4 8AA, United Kingdom.*

***Abstract*** — New types of tunable composite materials are considered, the effective microwave permittivity of which may depend on an external dc magnetic field or tensile stress. The composites consist of short pieces of conductive ferromagnetic wires embedded into a dielectric matrix. The short wire inclusions play a role of "the elementary scatterers", when the electromagnetic wave irradiates the composite and induces a longitudinal current distribution and electrical dipole moment in each inclusion. These induced dipole moments form the dipole response, which can be characterized by some complex effective permittivity. The later may have a resonance or relaxation dispersion caused by the strong current distribution along a wire, which depends on the wire high frequency surface impedance. In the vicinity of the resonance frequency any variations in the wire surface impedance result in a large change of the current distribution, and hence in the dipole moment of each inclusion and the effective permittivity on the whole. For a ferromagnetic conductive wire, the surface impedance may depend not only on its conductivity but also on the dc external magnetic field and tension through the so-called magneto-impedance effect (MI). Therefore, the dispersion of the effective permittivity can be tuned from a resonance type to a relaxation type, when a sufficient magnetic field or tensile stress is applied to the composite sample. A number of applications can be proposed, including the stress-sensitive media for remote non-destructive health monitoring of different structures, and selective microwave coatings with field-dependent reflection/transmission coefficients.

*)**Electronic mail:** dmakhnovskiy@plymouth.ac.uk

## I. Introduction and background research

Metal-dielectric composites demonstrate the fertile palette of electrodynamic properties concerned with their response to electromagnetic fields.[1-5] Here should be considered the frequency and spatial dispersions of the effective parameters, the effect of percolation and the associated field fluctuations that may result in non-linear wave (optical) and conductive (ac and dc) effects. The dispersion properties of the effective parameters of a composite are caused by both the initial dispersion in the material parameters of its components (inclusions and host material), and also the current distribution induced in them. The effect of percolation gives rise to the formation of long clusters of the conductive particles or metal fractions, when their concentration reaches a critical value named the percolation threshold. It could be said that the phase transition from the dielectric to the conductor takes place in the vicinity of the percolation threshold. At the quasistatic limit, when the wavelength inside the composite matrix is much larger than the characteristic size of inclusions and their concentration is less than the percolation threshold, the field spatial fluctuation can be neglected. In this case, the effective parameters can be determined by means of the mean field approximation (self-consistent), which takes into account the interparticle interaction. It is obvious that the only frequency dispersion can be accounted for such a way. Different approaches in the mean field theory vary in the so-called “mixing law”, which realizes the univocal correspondence between the composite effective parameters and the material parameters of its components. At that, some statistical information about the composite microstructure may taken into account.[6,7]

In the strict sense, the mean field approximation is no longer valid, when the inclusion size becomes comparable with the wavelength. In this case, the local field fluctuations may result in an enhancement of the non-linear effects related to the conductivity or light scattering.[1] Another interesting manifestation of non-quasistatic behavior is the resonance or relaxation (Debye) dispersion of the effective permittivity $\varepsilon_{eff}$. This may take place in a composite consisting of short pieces of conductive wires embedded into a dielectric matrix.[8.9] In this case, the dispersion of $\varepsilon_{eff}(\omega)$ is mainly caused by the strong distribution of induced current along a wire inclusion. The character of this distribution depends on the wire conductivity. The relaxation dispersion arises from wires with low conductivity, whereas the resonance dispersion requires a high conductivity. For the resonance dispersion, the imaginary part $\mathrm{Im}(\varepsilon_{eff}(\omega))$ of the effective permittivity has its sharp maximum (resonance absorption) in the vicinity of a resonance frequency $f_{res} = \omega_{res}/2\pi$. To the contrary, the real part $\mathrm{Re}(\varepsilon_{eff}(\omega))$ reaches its maximum value at a certain frequency before $f_{res}$, and the minimum value after $f_{res}$. Between these maximum and minimum values the dispersion function of $\mathrm{Re}(\varepsilon_{eff}(\omega))$ rapidly drops with increasing frequency, demonstrating the so-called anomalous dispersion. Furthermore, at a certain concentration of the highly conductive wires, $\mathrm{Re}(\varepsilon_{eff}(\omega))$ may drop below zero after $f_{res}$ and becomes negative. For wire inclusions with a low conductivity, the maximum of $\mathrm{Im}(\varepsilon_{eff}(\omega))$ becomes blurred and $\mathrm{Re}(\varepsilon_{eff}(\omega))$ decreases smoothly between its maximum and minimum values. The spectrum of resonance frequencies is almost defined by the wire length $l$ and the matrix permittivity $\varepsilon$: $f_{res,n} \sim c(2n-1)/\left(2l\sqrt{|\varepsilon|}\right)$, where $c$ is the velocity of light, and $n \geq 1$ is an integer. That gives us the right to designate $f_{res}$ as the antenna resonance. The dispersion amplitude at the first resonance ($n = 1$) appears much bigger than for subsequent ones. The attempts to generalize the mean field theory for such composites were undertaken in Refs. 8 and 10, where a scale dependent effective permittivity around the wire inclusions was introduced. A detailed experimental investigation of the dispersion properties of $\varepsilon_{eff}(\omega)$ carried out in Ref. 9 by the free-space technique has proved the availability of the effective parameters for these strongly non-quasistatic composites.

The effects discussed in our present work bear on a linear dipole response of the composite system irradiated by the electromagnetic fields. In the frame of this approach, the inclusions play the role of “the elementary scatterers” (or “atoms”), when an electromagnetic wave irradiates the composite and induces an electrical dipole moment in each inclusion. These induced dipole moments form the dipole response of the composite, which can be characterized by some effective permittivity. The use of metal inclusions is often justified by their larger polarisability in comparison with dielectric ones. This enables one to obtain a strong response even for very small inclusion concentrations. The shape of metal particles significantly determines their polarization properties. A small metal sphere or ellipsoid is the simplest shape, the scattering on which can be analyzed by means of Mie’s theory.[11,12] However, the most wide variety of polarization effects are observed in the wire inclusions with different spatial configurations, beginning at needle-shaped[1,5,8-10] and ending with spirals or omega-particles,[13,14] which may form artificial magnetism and chiral effective properties.[3,4]

The use of ferromagnetic wires as the inclusions in metal-dielectric composites significantly expands their functional capabilities. To our knowledge, the composites based on amorphous ferromagnetic wires were considered for the first time in Ref. 15. The samples were made of thin resin sheets with the thickness less than 2 mm filled with the short wire pieces with the diameter 3 μm and length 3 mm. An alloy of FeBSiMnC was used to fabricate wires with the positive magnetostriction resulting in the longitudinal magnetization. These wires demonstrate the ferromagnetic resonance (FMR) in the range $8-10$ GHz with the halfwidth $1-1.5$ GHz. The fabricated composite samples possess a radio-absorbing property with a resonance absorption of about of 30 dB within $8-10$ GHz and the halfwidth $1-1.5$ GHz. Since the antenna resonance $f_{res}$ $(n=1)$ is about $20-25$ GHz for $l=3$ mm and $\varepsilon \sim 4-6$ (resin permittivity), the observed resonance absorption should be attributed to FMR in the wire inclusions. The tunable properties were not investigated in that work. However, as it will be shown in our present work, the Fe-based wire inclusions with a longitudinal magnetization are not suitable to provide tunable properties, mostly due to the very large magnetic fields required to tune the ferromagnetic resonance frequency. Furthermore, it remains questionable why it was so important to use the magnetic wires instead of non-magnetic wires, which provides the same selective absorption but at the antenna resonance $f_{res}$.[9]

Tunable properties of the composites containing ferromagnetic wires were considered for the first time in Ref. 16. The high-frequency permittivity of composites consisting of a lattice of long ferromagnetic wires was considered. It was shown that the dielectric response is strongly dependent on the magnetic properties of the wires. A moderate external field induced large changes in the dielectric response. Also, negative real permittivity was observed over a wide frequency range for wires with circumferential magnetization, while a resonant behavior was observed on the axially magnetized wires. The subject of this work has been evoked by a large stream of publications on the so-called "metamaterials" or "left-handed" materials which all demonstrated the negative permittivity and permeability.[17] The negative permittivity in such materials at GHz frequencies can be introduced by the lattice of long wires, which have characteristic features of a metallic response to radiation. Contrary to composites with short inclusions, the electromagnetic field is applied locally to a certain portion of the material excluding the ends of wires. In this case the current distribution in the wire can be neglected. The most interesting results are obtained for wave polarization where the electric field is directed along to the wires. Such wire-mesh systems model the response of a diluted plasma,[1] giving a negative permittivity $\varepsilon_{eff}(\omega)$ below the normalized plasma frequency $\widetilde{\omega}_p = \omega_p / \sqrt{\varepsilon}$ somewhere in the gigahertz range: $\varepsilon_{eff}(\omega) = \varepsilon - \omega_p^2 / \omega^2$, where $\varepsilon$ is the matrix permittivity and $\omega_p$ is the "plasma frequency". In a general case, when the skin effect is not very strong, the plasma frequency depends on the wire impedance. In the case of ferromagnetic wires, their impedance depends on an external dc magnetic field – the so-called magneto-impedance. Therefore, the effective permittivity of such wire-mesh materials can be controlled by a dc magnetic field, as proposed in Ref. 16. The effect of magneto-impedance (MI), originally discovered in ferromagnetic wires,[18] has found important applications for magnetic sensors in the MHz range.[19-22] The ideas suggested in Ref. 16 have inspired some new applications for MI in the GHz range. Nevertheless, we have some serious reservations about the experimental results obtained in this work. First of all, the conclusions concerning the tunable properties of the effective permittivity were made from measurements of the reflection signal in the coaxial line loaded by only one wire. This is a nonsense because the effective parameters are employed to characterize the effective response of a composite system irradiated by free-space electromagnetic waves. Only the free-space technique can be responsible for verifying the tunable properties of the effective parameters. The problem was argued by the difficulties of achieving a homogeneous magnetic field through the sample. However, in the absence of a dc external magnetic field the effective permittivity was measured in free space for the composite samples made in real size. Besides, we are not sure that the authors were able to avoid the current distribution in the wire strings. In this case, the dispersion of effective permittivity will be formed by both the current distribution and initial dispersion of the wire material parameters. Therefore, we can conclude that the concept provided in this work prevails over the experimental results.

In our present work the effective permittivity is theoretically investigated for composite materials consisting of a dielectric matrix filled with short pieces of ferromagnetic wires. The wire inclusions in the composite matrix may have a random orientation or be ordered in some direction, depending on the application. It has been shown that for the Co-based amorphous wires having a circumferential or helical magnetization, the effective permittivity $\varepsilon_{eff}$ may strongly depend on the dc external magnetic field $H_{ex}$ or tensile stress $\sigma_{ex}$ applied to the composite sample as a whole. The consideration of these effects is based on the author's works, where field-tunable[23] and stress-tunable[24] composite materials have been proposed. Similar to the composites containing non-magnetic conductive wires, in the vicinity of $f_{res}$ the dispersion of $\varepsilon_{eff}(\omega)$ may have the resonance or relaxation type, depending of the wire impedance. Due to the MI effect, the wire impedance depends not only on its conductivity but also on an external dc magnetic field and tensile stress. Therefore, the dispersion of the effective permittivity can be tuned from a resonance type to a relaxation type, when the sufficient magnetic field or tensile stress is applied to the composite sample. The tensile stress is transmitted to each wire inclusion through the composite matrix. The magnetic fields that are required to tune the effective permittivity is about the same value as the anisotropy field $H_K$, which ranges $2-15$ Oe for the Co-based wires.

The content of the following two works is close to our subject, although they are not immediately connected with the effective material parameters of a composite irradiated by free space electromagnetic waves. A composite designed by Sensortex Inc (USA)[25] contains long thin Cu-wires that were fragmentary coated with a NiFe alloy with a high magnetic permeability and low coercitivity. These structured wires were embedded into a fiberglass/polyester material providing the mechanical strength. It was assumed to employ these composites as stress-sensitive materials. A separate Cu-wire was excited by a low frequency current (5 KHz) with a large amplitude (200–500 mA). The output voltages, measured across the fragments coated with the magnets, were strongly non-linear and demonstrated a high amplitude sensitivity to the local tensile stress transmitted through the composite matrix. A fill of such wires could image the pattern of the stress distribution over the composite sample. The used sensing-mechanism can not be attributed to the MI effect neither by the amplitude of excitation current nor its frequency. The MI assumes a linear response from a ferromagnetic sample subjected to a weak excitation (current or field) in the MHz or GHz ranges. To the contrary, a low frequency current with large amplitude will result in a magnetization reversal. In the considered case, the non-linear output voltage is proportional to the circular magnetic flux induced in the magnetic coating by an ac current flowing through the Cu-wire. In particular, this method is used for measurements of the circular B–H loops in magnetic wires, which can be obtained by the integration of the output voltage.[19] The magnetic flux can be changed by the external tensile stress through the inverse magnetostriction effect affecting the magnetic parameters.[26,27] The proposed stress monitoring method requires connections to the readout electronics.

In Ref. 28 the electrically conductive composites have been developed with self-diagnosis stress properties. For memorizing strain, carbon fibers or particles were added into the plastic composite matrix. A pre-tensile stress in the fiber-filled composite enhances its residual effective resistance that enables the detection of the smallest strains. The composite filled with the high fraction of carbon particles, which constitute a percolation conductive structure, demonstrates the effective residual resistance without application of a pre-tensile stress. The irreversible changes of the effective resistance of the composite depend on the strain history: the effective resistance increases in proportion to the logarithm of the number of tensile cycles. This stress-monitoring method also requires a contact measurement. Nevertheless, since the indicator parameter is the effective conductivity, this material can be characterized as a sensing medium.

## II. Typical ferromagnetic wires

The glassy alloys used for the fabrication of amorphous wires with soft ferromagnetic properties can be classified into two groups: Fe– and Co–based.[29-35] Owing to their resistance to crystallization, these easy glass-forming alloys can be cast in bulk shape with very small dimensions down to microns. The general composition of alloy is $F_xM_y$ with Fe and/or Co as "F" and metalloid like Si and B as "M". The content "$x$" ranges typically between 70 and 80%. The alloys may also contain small amounts of other elements such as Cr, Mn, Al, Cu, and Nb in order to improve mechanical, corrosion or magnetic properties.[36-39]

The magnetostriction plays the main role in determining the magnetic behavior (i.e. domain structure and hysteresis loop). The sign and value of the magnetostriction, $\lambda$, are decisive. Positive and negative magnetostrictions result in a radial and circumferential easy axis in the shell respectively, whereas the inner core always has longitudinal magnetization (although it can be very small). Thin amorphous ferromagnetic wires having a negative magnetostriction can be considered to be one of the best materials for MI applications, including tunable microwave composites. The wire sample with circumferential anisotropy ($\lambda < 0$) is divided into a "bamboo-like" domain structure, where adjacent domains have opposite directions of magnetization, as shown in Fig. 1. The magnetostriction constant $\lambda$ is correlated to the alloy composition. In the $(Co_{1-x}Fe_x)_{75}Si_{15}B_{10}$ series, $\lambda$ is positive for $x > 0.06$, and it becomes negative when the Fe-content drops below down this value.[33] Therefore, the negative magnetostriction is typical for Co-rich alloys. For example, a wire of a composition $Fe_{4.35}Co_{68.15}Si_{12.5}B_{15}$ exhibits excellent soft magnetic properties having almost zero (but still negative) magnetostriction of $\lambda \sim -10^{-7}$. The Co-rich wires with $\lambda < 0$ present an almost nonhysteretic magnetization curve. In contrast, Fe-rich wires with $\lambda > 0$ are characterized by a large Barkhausen jump that results in square-shaped hysteresis loops.

Currently, there are two main techniques of wire fabrication. Amorphous wires are made by UNITIKA LTD R&D (www.unitika.co.jp) using the in-water spinning method, and then cold drawn from a diameter of about 125 μm of the as-cast wire to diameters of 20–30 μm. The final sample undergoes annealing with a tension stress to build up a certain magnetic structure. This method requires very careful control of the annealing process to obtain repeatable magnetic parameters. Some commercial companies and research laboratories fabricate amorphous wires with a glass coating, by a modified Taylor-Ulitovskiy method.[40] For instance, Tamag Iberica S.L. (www.tamagiberica.com), GMWT LTD (www.gmwt-gw.com), ELIRI Institute (http://eliri.md/eng/), and MFTI LTD (www.microwires.com) produce a broad assortment of amorphous glass-coated wires (magnetic and non-magnetic) of the best quality and different compositions providing a wide range of magnetic properties and conductivities. During the fabrication process, one end of the glass tube with the alloy of the required composition is sealed. Then, it is heated to the temperature at which the glass is soften and the alloy is in a melting state. Drawing the heated end creates a very thin glass capillary (ranging between 10–60 μm) where the molten metal streams. The final wire structure is formed by the water cooling to obtain a metallic core (in amorphous state). The metallic core (ranging between 3–50 microns) has an amorphous and/or microcrystalline microstructure in order to achieve the desired magnetic properties, such as magnetic anisotropy and coercitivity. The ratio of the metal core and glass coating thickness also affects the magnetic properties. The fabrication method of glass-covered wires introduces a large internal tensile stress mainly arising from the difference in thermal expansion coefficients of metal (nucleus) and glass (sheath). The value of this internal stress, can be controlled by the ratio of diameters of the glass cover and the metal core. The direction of the easy magnetoelastic anisotropy axis associated with the stress distribution inside a wire is determined by the sign of the magnetostriction constant $\lambda$.[41-46] In context of cost and simplicity of the technological process, the glass-covered wires have a great advantage over those produced by the in-water spinning method.

### III. Effective permittivity in composites filled with the short conductive wires

As it has been mentioned in the Introduction, that the dispersion of $\varepsilon_{eff}$ has different origins for "short" and "long" inclusions. In the first case, the composites demonstrate the Lorentz dipole dispersion, whereas the second type of materials is characterized by the Drude dispersion of free-electron gas. The Lorentz model of dispersion is applicable to insulator materials. The composite with short inclusions is similar in many respects to an isolator since the wire-inclusions play a role of "atoms" (elementary dipole scatterers), which are polarized with an ac electric field. The wire inclusions in the composite matrix may have a random orientation or can be ordered in some direction, depending on the application. The local electrical field $e_{loc}\exp(-i\omega t)$ induces the current with a linear density $j(x)\exp(-i\omega t)$ distributed along the inclusion length. The electric dipole moment $D$ and the dielectric polarisability $\alpha$ of the inclusion are calculated using the continuity equation $\partial j(x)/\partial x = i\omega\rho(x)$ and integrating by parts with boundary conditions $j(\pm l/2) \equiv 0$ (here $\rho$ is the charge density per unit length):

$$D = \frac{i}{\omega}\int_{-l/2}^{l/2} j(x)dx \Rightarrow \alpha = D/(Ve_{loc}), \tag{1}$$

where $V$ is the inclusion volume. As it will be shown in next Section, the density $j(x)$ of a linear current can be approximated by a linear differential equation of the second order with the boundary conditions $j(\pm l/2) \equiv 0$. Thus, as in the case of a Lorentz oscillator the dispersion of the polarisability $\alpha$ has the following form:[47]

$$\alpha(\omega) = \sum_n \frac{A_n}{\left(\omega_{res,n}^2 - \omega^2\right) - i\Gamma_n\omega}, \tag{2}$$

where the summation is carried out over all antenna resonance frequencies $\omega_{res,n} = 2\pi f_{res,n} = 2\pi c/\lambda_{res,n}$ in increasing order, $\lambda_{res,n}$ are the resonance wavelengths, $A_n$ are some amplitude constants, $\Gamma_n$ are the dumping parameters. The first resonance ($n=1$) with the lowest frequency has a maximum amplitude $A_1$ and gives the main contribution to the polarisability. Each $\Gamma_n$ can be decomposed into two parts $\Gamma_n^{rad}$ and $\Gamma_n^{mr}$ related to the radiation and internal (magnetic and resistive) losses, respectively. The damping parameter $\Gamma_n^{mr}$ involving magnetic losses may depend on an external magnetic field $H_{ex}$ or tensile stress $\sigma_{ex}$. In the vicinity of an antenna resonance the polarisability $\alpha$ will strongly depend on $H_{ex}$ or $\sigma_{ex}$ if the condition $\Gamma_n^{mr} \sim \Gamma_n^{rad}$ is held.

The bulk polarization **P** of the composite is of the form: $\mathbf{P} = < e_{loc} > p\boldsymbol{\alpha} = \mathbf{e}_0 \vartheta_{eff}$, where $< e_{loc} >$ is the averaged local field, $p$ is the volume concentration of the inclusions, $\mathbf{e}_0$ is the external electrical field, and $\vartheta_{eff}$ is the effective bulk susceptibility. In the vicinity of $f_{res}$ the effective response of composite has a strongly non-quasistatic dispersion, as seen in Eq. (2). Nevertheless, it is still possible to introduce the effective susceptibility $\vartheta_{eff}$. First of all, the wavelength of the incident and reflected electromagnetic waves outside of the composite matrix with a permittivity $\varepsilon >> 1$ appear much larger than the wire length: $\lambda_{out} \sim 2l\sqrt{\varepsilon}$ (in the vicinity of the first resonance). Therefore, the scattered wave will have a dipole character not far from the sample surface. Thus, the non-quasistatic behavior concerns only the interaction between wires inside the composite matrix, where $\lambda_{in} = \lambda_{out}/\sqrt{\varepsilon} \sim 2l$ that results in the strong field fluctuations. An attempt to generalize the mean field theory (MFT) for such composites was undertaken in Ref. 8, where the scale dependence of the effective permittivity around the wires was introduced. This approach was employed in Ref. 10 to derive the MFT equation for a thin composite sample taking into account the boundary effects on the wire polarization. But the problem remains open because the use of MFT in the non-quasistatic case is very questionable. Moreover, the strong current distribution along the wire inclusions may result in a spatial dispersion of the effective permittivity (see discussion in Section V).

To relate the polarisability $\boldsymbol{\alpha}$ to the effective bulk susceptibility $\vartheta_{eff}$, the relation between $e_{loc}$ and $e_0$ has to be established. For small inclusion concentrations $p << p_c$, where $p_c$ is the percolation threshold, it is reasonable to assume that $< e_{loc} > \approx e_0$, which leads to:

$$\varepsilon_{eff} \approx \varepsilon + 4\pi\, p < \boldsymbol{\alpha} >, \tag{3}$$

where $\varepsilon$ is the matrix permittivity (complex in general), $< \boldsymbol{\alpha} >$ is the polarisability of an individual inclusion averaged over its orientations. Any mean field theories will result in this single particle approximation in the limit of a low inclusion concentration. For a composite filled with the wire inclusions the percolation threshold occurs at a very low concentration: $p_c \sim 2a/l$, where $a$ is the wire radius (microns).[8] Nevertheless, an even lower concentration will give rise the strong dipole response due to the large polarization of the elongated inclusions. The single particle approximation enables the qualitative explanation to all of the effects observed in the wire composites, including the resonance or relaxation dispersion, and tunable properties.

The overwhelming majority of applications using composite materials require that a composite sample be prepared as a thin layer or as an additional surface cover. A composite layer with a random orientation of wire inclusions is shown in Fig. 2. For this composite structure we can accept that $< \boldsymbol{\alpha} > = \alpha / 2$, where the polarisability $\alpha$ has to be calculated from Eq. (1) for the known current distribution $j(x)$. This distribution can be found from the antenna equation described in next Section.

## IV. Impedance tensor and the generalized antenna equation

Within the framework of a single particle approximation the scattering problem for a thin wire has to be solved. As it will be shown below, the electromagnetic response from a thin conductive wire can be described by means of the effective linear current flowing along the axis and having only an axial distribution. Within the framework of this approach, known as the antenna approximation,[48] the wavelength and the wire length are assumed to be much larger than the wire cross size $2a$. For a start, let us consider the incident wave having a longitudinal electric field $\bar{e}_{x0}$ at the surface of a non-magnetic wire ($x$ is the coordinate along the wire). In this case, the total induced current is longitudinal, which determines the scattered electromagnetic field having longitudinal electric $\bar{e}_x$ and circular magnetic $\bar{h}_\varphi$ components at the wire surface, where $\varphi$ is the azimuthal coordinate. The same polarization $(\bar{e}_x, \bar{h}_\varphi)$ can be induced by a linear current with the volume density $j(x)\delta_S$ flowing along the axis, where $x$ is a point on the axis and $\delta_S$ is the two dimensional Dirac's function. Further, the function $j(x)$ will be referred to as "linear density" or "density". Thus, the linear longitudinal current plays a role of an effective current producing the surface field of the required polarization $(\bar{e}_x, \bar{h}_\varphi)$ and intensity. If the incident electromagnetic field contains a longitudinal magnetic component $\bar{h}_{x0}$ at the wire surface, a circular electric field $\bar{e}_\varphi$ will be induced in a wire (magnetic or non-magnetic). In this case, a longitudinal linear current does not provide the total polarization of the scattered field. Furthermore, for a magnetic wire, the field $\bar{h}_{x0}$ will induce $\bar{e}_x$, and the field $\bar{e}_{x0}$ will induce $\bar{h}_\varphi$. However, the total scattered field can be decomposed into two basic waves with polarizations at the wire surface: $(\bar{e}_x, \bar{h}_\varphi)$ and $(\bar{e}_\varphi, \bar{h}_x)$, where $(\bar{e}_x, \bar{h}_\varphi)$ is determined by the linear current. The other polarization $(\bar{e}_\varphi, \bar{h}_x)$ can be calculated directly from the impedance boundary condition, which represents a linear relationship between $\bar{\mathbf{e}}$ and $\bar{\mathbf{h}}$ on the wire surface (see below). Nevertheless, the polarization effects arising due to the radial electric and magnetic fields can be neglected because of their small dipole moments in comparison with the dipole moment induced by the longitudinal current. Thus, the concept of the linear effective current describes correctly the scattered field at any polarization of the incident wave.

The response from a thin wire irradiated by an electromagnetic field can be fully determined from the external scattering problem with the boundary conditions at the wire surface, which are set via the surface impedance tensor $\hat{\varsigma}$:[49,50]

$$\overline{\mathbf{E}}_t = \hat{\varsigma}[\overline{\mathbf{H}}_t \times \mathbf{n}], \qquad (4)$$

where the square brackets designate the vector multiplication, $\mathbf{n}$ is the unit normal vector directed inside the wire, $\overline{\mathbf{E}}_t$ and $\overline{\mathbf{H}}_t$ are the tangential vectors of the total electric and magnetic fields at the wire surface, which include both the scattered and excitation fields. In Eq. (4) we assume that the impedance tensor $\hat{\varsigma}$ is the local characteristic of a conductive sample. In the case of an ideally conductive wire (the conductivity $\sigma = \infty$), the condition (4) nulls: $\overline{\mathbf{E}}_t \equiv 0$, which is typically used in the antenna problems.

Boundary condition (4) is convenient to write in the local cylindrical co-ordinate system $(x, \varphi, r)$ related to the wire:

$$\begin{aligned} \overline{E}_x &= \varsigma_{xx}\overline{H}_\varphi - \varsigma_{x\varphi}\overline{H}_x \\ \overline{E}_\varphi &= \varsigma_{\varphi x}\overline{H}_\varphi - \varsigma_{\varphi\varphi}\overline{H}_x \end{aligned}. \qquad (5)$$

Within the antenna approximation, the field $\overline{H}_\varphi(x)$ contains only the circular field $\bar{h}_\varphi(x)$ induced by a current with linear density $j(x)$. On the contrary, the longitudinal field $\overline{H}_x$ is entirely defined by the excitation field $\bar{h}_{0x}$. For a non-magnetic wire, the off-diagonal terms $\varsigma_{x\varphi} = \varsigma_{\varphi x} \equiv 0$. The total magneto-impedance tensor $\hat{\varsigma}$ was found in Ref. 50 for a ferromagnetic wire with an arbitrary type of the magnetic anisotropy for any frequency. For wires with circumferential and helical anisotropies, $\hat{\varsigma}$ was measured over the MHz frequency range in Ref. 51 and 52, respectively.

The generalized antenna equation for $j(x)$ in a thin wire with the impedance boundary conditions (5) was derived in Ref. 23:

$$\frac{\partial^2}{\partial x^2}(G*j)+k^2(G*j)=\frac{i\omega\varepsilon}{4\pi}\bar{e}_{0x}(x)-\frac{i\omega\varepsilon\varsigma_{xx}}{2\pi\, ac}(G_\varphi*j)+\frac{i\omega\varepsilon\varsigma_{x\varphi}}{4\pi}\bar{h}_{0x}(x)\,. \quad (6)$$

Here, it is assumed that the wire is embedded into the matrix with the permittivity $\varepsilon$ and permeability $\mu$ (both the complex, in general), $k=(\omega/c)\sqrt{\varepsilon\mu}$ is the wave number in the dielectric matrix, and the operator in the brackets designates the convolution of a function with $j(x)$:

$$(G*j)=\int_{-l/2}^{l/2}G(x-s)j(s)ds\,,\quad (G_\varphi*j)=\int_{-l/2}^{l/2}G_\varphi(x-s)j(s)ds\,. \quad (7)$$

The function $G$ is the Green function of the scalar Helmholtz equation:

$$G(r)=\frac{\exp(ikr)}{4\pi\, r}\,, \quad (8)$$

where $r=\sqrt{(x-s)^2+a^2}$ .

The function $G_\varphi(r)$ determines the scattered circular magnetic field $\bar{h}_\varphi(x,a)$ at the wire surface:

$$\bar{h}_\varphi(x,a)=\frac{2}{ac}(G_\varphi*j)=\frac{2}{ac}\int_{-l/2}^{l/2}G_\varphi(r)j(s)ds\,, \quad (9)$$

$$G_\varphi(r)=\frac{a^2(1-ikr)\exp(ikr)}{2r^3}\,.$$

Equation (6) has to be completed imposing the boundary conditions at the ends of the conductor:

$$j(-l/2) = j(l/2) \equiv 0\,. \tag{10}$$

The integro-differential equation (6) involves general losses including both the radiation and internal losses (resistive losses and magnetic relaxation). The internal losses appear via the impedance component $\varsigma_{xx}$ and the convolution $(G_\varphi * j)$, whereas the imaginary part of $(G * j)$ determines the radiation losses. Along with this, there is an additional term in the right part of Eq. (6), related with the off-diagonal impedance component $\varsigma_{x\varphi}$. In the ferromagnetic wires with a circumferential anisotropy and "bamboo-like" domain structure shown in Fig. 1 the averaged off-diagonal components $\varsigma_{x\varphi}$ and $\varsigma_{\varphi x}$ are almost zero, as was proved theoretically[50] and experimentally.[51] The effects related to the off-diagonal components are possible only in a wire with a helical anisotropy where such averaging does not occur: $\varsigma_{x\varphi} \neq 0$.[50,52] Composites containing such "exotic" wire inclusions may exhibit chiral properties.[53] This is discussed in Section V.

As it follows from Eqs. (8) and (9), the real functions $\mathrm{Re}(G)$ and $\mathrm{Re}(G_\varphi)$, considered at the wire surface, have a sharp peak at $r = a$. Thus, $\mathrm{Re}(G)$ and $\mathrm{Re}(G_\varphi)$ give the main contribution to Eq. (6): $|(\mathrm{Im}(G) * j)| << |(\mathrm{Re}(G) * j)|$ and $|(\mathrm{Im}(G_\varphi) * j)| << |(\mathrm{Re}(G_\varphi) * j)|$. However, the convolutions with the imaginary parts are important in the vicinity of the resonance and can be taken into account by an iteration procedure, which was developed in Refs. 23 and 54. For the calculation of convolutions with functions $\mathrm{Re}(G)$ and $\mathrm{Re}(G_\varphi)$ it is possible to use an approximate method:[10]

$$(\mathrm{Re}(G) * j) \approx j(x) \int_{-l/2}^{l/2} \mathrm{Re}(G(r))ds = j(x)Q\,, \tag{11}$$

$$(\mathrm{Re}(G_\varphi)*j) \approx j(x)\int_{-l/2}^{l/2}\mathrm{Re}(G_\varphi(r))ds = j(x)Q_\varphi\,.$$

Here:

$$Q = \int_{-l/2}^{l/2}\mathrm{Re}(G(r))ds \sim \frac{1}{4\pi}\int_{-l/2}^{l/2}\frac{ds}{\sqrt{s^2+a^2}} \sim \frac{\ln(l/a)}{2\pi}\,, \tag{12}$$

$$Q_\varphi = \int_{-l/2}^{l/2}\mathrm{Re}(G_\varphi(r))ds \sim \frac{a^2}{2}\int_{-l/2}^{l/2}\frac{ds}{(s^2+a^2)^{3/2}} + \frac{a^2k^2}{2}\int_{-l/2}^{l/2}\frac{ds}{\sqrt{s^2+a^2}} \sim \left(1+a^2k^2\ln(l/a)\right),$$

where $Q$ and $Q_\varphi$ are the positive form-factors.

From Eq. (6) and inequalities $|(\mathrm{Im}(G)*j)| << |(\mathrm{Re}(G)*j)|$ and $|(\mathrm{Im}(G_\varphi)*j)| << |(\mathrm{Re}(G_\varphi)*j)|$, we obtain differential equation for the zero approximation $j_0(x)$ where the radiation losses are neglected:

$$\frac{\partial^2}{\partial x^2}j_0(x) + \left(\frac{\omega}{c}\right)^2\varepsilon\mu\left(1+\frac{ic\varsigma_{xx}}{2\pi\, a\omega\,\mu}\frac{Q_\varphi}{Q}\right)j_0(x) \approx \frac{i\omega\,\varepsilon}{4\pi\, Q}\left(\bar{e}_{0x}(x)+\varsigma_{x\varphi}\,\bar{h}_{0x}(x)\right). \tag{13}$$

As it follows from Eq. (13), the implementation of the impedance boundary condition leads to the renormalization of the wave number, which becomes:[23]

$$\widetilde{k} = \frac{\omega}{c}\sqrt{\varepsilon\mu}\left(1+\frac{ic\varsigma_{xx}}{2\pi\, a\omega\,\mu}\frac{Q_\varphi}{Q}\right)^{1/2}. \tag{14}$$

The effective wave number $\widetilde{k}$ defines the resonance wavelengths and frequencies of Eq. (13) ($k_{res}l = \pi(2n-1)$):

$$\lambda_{res,n} = \frac{2l\sqrt{\varepsilon\mu}}{2n-1}\sqrt{\mathrm{Re}\left(1+\frac{ic\varsigma_{xx}}{2\pi\, a\omega\,\mu}\frac{Q_\varphi}{Q}\right)},\ n = 1,2,3... \tag{15}$$

Therefore, $f_{res,n} = c/\lambda_{res,n}$ is the spectrum of resonance frequencies.

The values in Eq. (15) differ from the resonance wave lengths obtained for an ideally conductive wire: $\lambda_{res,n} = 2l\sqrt{\varepsilon\mu}/(2n-1)$. The effect of a limited conductivity will result in a shifting of $f_{res,n}$ towards higher frequencies.

The equation (6) was written for a single wire. In strict sense, the interactions between wires in a composite structure can be taken into account when they are ordered in the same direction and placed in the points of some periodical lattice. Such a composite system can be named a "wire crystal",[1,5,55-58] where the current distribution in each wire will be the same due to the symmetry. To calculate this current distribution, the Green function $G$ in Eq. (6) has to be replaced with the lattice Green function $\hat{G}$,[59] which is the sum of $G$ over the lattice:

$$\hat{G}(x-s,y,z) = \sum_{q,l,m} \frac{\exp(ik\, r_{q,l,m})}{4\pi\, r_{q,l,m}}, \tag{16}$$

where, $r_{q,l,m} = \sqrt{(x-s-x_q)^2 + (y-y_l)^2 + (z-z_m)^2}$ and the point $(x,y,z)$ lies onto the wire surface, and $(q,l,m)$ is the 3D integer index of a wire in the wire lattice. The function $\hat{G}$ is the lattice invariant. The vector $(x_q, y_l, z_m)$ is directed to a wire centre. With $q,l,m \equiv 0$, $r_{0,0,0} = \sqrt{(x-s)^2 + a^2}$. With $l \neq 0$ and $m \neq 0$, $r_{q,l,m} \approx \sqrt{(x-s-x_q)^2 + (y_l)^2 + (z_m)^2}$, since $a$ is assumed to be much smaller, than the lattice constants. The onrush of nano-technology makes possible the fabrication of wire particles of different shapes even at nano-scales, including nano-wires and nano-rings which can arranged into a periodical lattice.[60,61] The universality of the generalized antenna equation (6) has allowed us to calculate the scattering in nano-wire-particles at optical frequencies.[54]

When a wire is embedded into a very thin dielectric layer, Eq. (6) will take some different form:

$$\frac{\partial^2}{\partial x^2}\left((\widetilde{G}+\widetilde{U})*j\right)+k_2^2(\widetilde{G}*j)=\frac{i\omega\varepsilon}{4\pi}\bar{e}_{0x}(x)-\frac{i\omega\varepsilon\varsigma_{xx}}{2\pi\, ac}(G_\varphi*j)+\frac{i\omega\varepsilon\varsigma_{x\varphi}}{4\pi}\bar{h}_{0x}(x), \quad (17)$$

$j(-l/2)=j(l/2)\equiv 0$.

Here, the function $\widetilde{G}$ and $\widetilde{U}$ take into account the effect of boundaries:[10]

$$\widetilde{G}(r)=\frac{\exp(ik_2 r)}{4\pi r}+\left(\frac{1}{4\pi}\int_0^{+\infty}\frac{\mathrm{a}_{x2}(k,h,h_0)\exp(-\gamma_2 h_0)+\mathrm{b}_{x2}(k,h,h_0)\exp(\gamma_2 h_0)}{\Delta_2(k,h)\gamma_2}\mathrm{J}_0(k\rho)k\,dk\right)^*, \quad (18)$$

$$\widetilde{U}(r)=\left(\frac{1}{2\pi}\int_0^{+\infty}\frac{\left(\mathrm{a}_{z2}(k,h,h_0)\exp(-\gamma_2 h_0)-\mathrm{b}_{z2}(k,h,h_0)\exp(\gamma_2 h_0)\right)(\varepsilon\mu-1)}{\Delta_1(k,h)\Delta_2(k,h)}\gamma_2\,\mathrm{J}_0(k\rho)k\,dk\right)^*, \quad (19)$$

where "*" is complex conjugation (in Ref. 10 the different time dependence $\exp(+i\omega t)$ was used), $h$ is the dielectric layer thickness, $h_0$ is the depth of immersion of the wire into the dielectric layer, $r=\sqrt{(x-s)^2+a^2}$, $\rho=\sqrt{x^2+a^2}$, $\gamma_1(k)=\sqrt{k^2-k_1^2}$, $\gamma_2(k)=\sqrt{k^2-k_2^2}$, $k_1=\omega/c$ and $k_2=(\omega/c)\sqrt{\varepsilon\mu}$ are the wave numbers in free space and dielectric layer, and $\mathrm{J}_0$ is the Bessel function. Integrals in Eqs. (18) and (19) use the following functions:

$$\Delta_1(k,h)=\left(\gamma_2^2+\gamma_1^2\varepsilon^2\right)\mathrm{sh}(\gamma_2 h)+2\gamma_1\gamma_2\varepsilon\,\mathrm{ch}(\gamma_2 h),$$

$$\Delta_2(k,h)=\left(\gamma_2^2+\gamma_1^2\mu^2\right)\mathrm{sh}(\gamma_2 h)+2\gamma_1\gamma_2\mu\,\mathrm{ch}(\gamma_2 h),$$

$$\mathrm{a}_{x2}=(\gamma_2-\gamma_1\mu)\left[\gamma_1\mu\,\mathrm{sh}(\gamma_2(h-h_0))+\gamma_2\,\mathrm{ch}(\gamma_2(h-h_0))\right],$$

$$\mathrm{b}_{x2}=\exp(-\gamma_2 h)(\gamma_2-\gamma_1\mu)\left[\gamma_1\mu\,\mathrm{sh}(\gamma_2 h_0)+\gamma_2\,\mathrm{ch}(\gamma_2 h_0)\right],$$

$$\begin{aligned}\mathrm{a}_{z2}=&\left[\gamma_1\mu\,\mathrm{sh}(\gamma_2(h-h_0))+\gamma_2\,\mathrm{ch}(\gamma_2(h-h_0))\right](\gamma_2+\gamma_1\varepsilon)\exp(\gamma_2 h)-\\&-\left[\gamma_1\mu\,\mathrm{sh}(\gamma_2 h_0)+\gamma_2\,\mathrm{ch}(\gamma_2 h_0)\right](\gamma_2-\gamma_1\varepsilon)\end{aligned},$$

$$\begin{aligned}\mathrm{b}_{z2}=&\left[\gamma_1\mu\,\mathrm{sh}(\gamma_2(h-h_0))+\gamma_2\,\mathrm{ch}(\gamma_2(h-h_0))\right](\gamma_2-\gamma_1\varepsilon)\exp(-\gamma_2 h)-\\&-\left[\gamma_1\mu\,\mathrm{sh}(\gamma_2 h_0)+\gamma_2\,\mathrm{ch}(\gamma_2 h_0)\right](\gamma_2+\gamma_1\varepsilon)\end{aligned}.$$

Using Eqs. (18) and (19), the following limits for the functions $\widetilde{G}(r)$ and $\widetilde{U}(r)$ can be obtained in two limiting cases $h \to 0$ and $h \to \infty$:

$$\lim_{h\to 0}\widetilde{U} = \frac{(\varepsilon\mu - 1)}{\mu}\frac{e^{ik_1 r}}{4\pi r}, \qquad \lim_{h\to\infty}\widetilde{U} = 0, \tag{20}$$

$$\lim_{h\to 0}\widetilde{G} = \frac{1}{\mu}\frac{e^{ik_1 r}}{4\pi r}, \qquad \lim_{h\to\infty}\widetilde{G} = \frac{e^{ik_2 r}}{4\pi r}.$$

The combined effect of the layer boundaries and impedance conditions (5) will result in a very complicated self-consistent dispersion equation for the resonance wavelengths:

$$\lambda_{res,n} = \frac{2l\sqrt{\varepsilon\mu}}{2n-1}\sqrt{\mathrm{Re}\left(\frac{Q_2(\lambda_{res,n},h,h_0)}{Q_1(\lambda_{res,n},h,h_0)} + \frac{ic\varsigma_{xx}}{2\pi a\omega\mu}\frac{Q_\varphi(\lambda_{res,n},h,h_0)}{Q_1(\lambda_{res,n},h,h_0)}\right)};\ n = 1,2,3... \tag{21}$$

By analogy with Eq. (12) the form factors $Q_1$ and $Q_2$ are calculated from the following integrals:

$$Q_1 = \int_{-l/2}^{l/2}\mathrm{Re}(\widetilde{G}(r) + \widetilde{U}(r))ds, \tag{22}$$

$$Q_2 = \int_{-l/2}^{l/2}\mathrm{Re}(\widetilde{G})ds.$$

The thickness dependence of the resonance frequency $(n = 1)$ has been investigated theoretically and experimentally.[10,62,63] The resonance frequency changes from the vacuum value at $h \to 0$ (defined by Eq. (15) with $\varepsilon = \mu \equiv 1$) to the value corresponding to an infinite medium. The characteristic feature of this thickness dependence is the existence of two regions defined by the critical thickness $h_c$. For $h < h_c$, the resonance frequency rapidly drops with increasing $h$, and for $h > h_c$ it decreases slowly reaching the saturation limit. It was shown that $h_c$ is independent of the material parameters of the dielectric layer, and it is determined by only the wire length. For a wire with length 1 cm, $h_c$ was found to be $\sim 200$ μm.

It would be useful to establish the role of the surface waves in the forming of the effective response in a thin composite layer. The main method used in Ref. 10 is the solution of the heterogeneous Helmholtz equation $\Delta u + k^2 u = j(\mathbf{r})$ in the layered structure. To extract an unequal solution of this equation in an infinite region, which is the exterior of a finite region, it is needed to assume the additional limitations on the behavior of the solution when approaching infinity. These additional limitations are the famous Sommerfield conditions:

$$\frac{\partial u(\mathbf{r})}{\partial |\mathbf{r}|} \pm iku(\mathbf{r}) = o(1/|\mathbf{r}|), \quad |\mathbf{r}| \to \infty, \tag{23}$$

where the signs "$\pm$" correspond to the outgoing and arriving waves. In the strict sense, these conditions are required only if the exterior region does not have any energy losses. To the contrary, if the exterior region has the losses, then the amplitude of the outgoing wave (scattered) must decrease when approaching infinity, whereas for the arriving wave it must increase. However, it is a little known fact that the Sommerfield conditions must be generalized in the case of the layered media. Conditions (23) contain the roots $\pm k$ of the simplest dispersion equation $p^2 - k^2 = 0$, which defines the poles of the Fourier transform of the Green function corresponding to the Helmholtz equation in free space (see Eq. (8)). In a layered structure, the equation of poles will be very complicated and the radiation conditions (23) must be formulated for each root of this dispersion equation.

The Fourier images of the Green functions in the dielectric layer contain the poles, which are found from the dispersion equations $\Delta_1 = 0$ and $\Delta_2 = 0$ (see Eqs. (18) and (19)). As it was mentioned above these poles relate to the spectrum of the surface waves, which lies between two wave numbers $k_1$ (free space) and $k_2$ (dielectric layer). With increasing layer thickness this spectrum tends to be dense everywhere, covering the interval $[k_1, k_2]$, i.e. it becomes continuous. In this case, the dielectric layer demonstrates the so-called "soft dynamic properties", when it allows the propagation of the waves scattered on an embedded wire inclusion with any wave number between $k_1$ and $k_2$. Thus, from the point of view of this wire inclusion (an opened resonator) a thick dielectric layer can be priory characterized by some permittivity. When decreasing the layer thickness, the spectrum of the allowed surface waves significantly converges. In this case, the effective properties of the dielectric layer can not be characterized by an effective permittivity for an embedded wire inclusion, since this permittivity requires a certain wave number which may lie out of the allowed spectrum. Nevertheless, there is a compromise, when the embedded wire inclusion tries to agree with the dielectric layer about a possible surface wave, which can propagate. As a result of this compromise, the certain wave number arises, with which the scattered wave leaves the wire inclusion. This new wave number can be found from the "compromise dispersion equation" (21). Thus, we can conclude that the effective response of a thin composite system is determined by the spectrum of elementary excitations allowed by the dielectric layer. This conclusion is very unusual for the quasistatic insight, which has got into the habit of located parameters (capacitance, resistance and inductance). But we are sure that wave processes are determinative for thin composite systems.

### IV. Field and stress dependences of the magneto-impedance

At the present, the magneto-impedance effect (MI) is treated as a generalized Ohm's law, where the magneto-impedance takes the non-diagonal tensor form (5). All the types of linear excitations and responses in a ferromagnetic sample, including wires[50] and thin film,[64,65] can be expressed in the terms of the impedance matrix components. The MI has an electrodynamic origin owing to the redistribution of the ac current density under the application of the dc magnetic field. In the original theoretical works on MI[18] the current density has been calculated with the assumption that the variable magnetic properties can be described in terms of a total permeability having a scalar or quasi-diagonal form. This allows the impedance of a magnetic sample (and the voltage induced across it by an ac current $j\exp(-i\omega t)$) to be found essentially in the same way as in the case of a non-magnetic material.[49] In this approach, the voltage response $V$ measured across the sample is of the form $V = Z(a/\delta_m)j$, where the impedance $Z$ is calculated as a function of the magnetic skin depth $\delta_m = c/\sqrt{2\pi\sigma\omega\mu_t}$. Here $\sigma$ is the sample conductivity and $\mu_t$ is the effective transverse permeability (with respect to the current flow). If the skin effect is strong ($\delta_m/a << 1$), the impedance is inversely proportional to the skin depth, therefore, the magnetic-field dependence of the transverse permeability controls the voltage behavior. This simple consideration has provided a qualitative understanding of the MI behavior, and in certain cases it gives a reasonable agreement with the experimental results. A good example is the MI effect in a Co-based amorphous wire. A tensile stress from quenching (and enhanced by tension annealing) coupled with the negative magnetostriction ($\lambda < 0$) results in a circumferential anisotropy and a corresponding left and right handed alternative circular domain structure (see Fig. 1). The ac current passing through the wire induces an easy-axis magnetic field which moves the circular domain walls so that they nearly cross the entire wire. The circular magnetization is very sensitive to the axial dc magnetic field which is a hard axis field. The ac permeability associated with this process is circumferential and corresponds to $\mu_t$ introduced above. Substituting in $\delta_m = c/\sqrt{2\pi\sigma\omega\mu_t}$ this circular permeability accounting for the field dependence and the frequency dispersion due to the local domain wall damping gives a very good agreement with the experimental MI spectra for frequencies lower than the characteristic frequency of the domain wall relaxation ($\sim 1-10$

MHz for 30 micron diameter wires).[18] Typically, the rotational relaxation is a faster process, and for higher frequencies ($>10$ MHz) the magnetization rotation dynamics dominates. The rotational permeability has an essential tensor form, which makes it difficult to use scalar equations for higher frequencies: the difference between the experimental and theoretical results becomes quite considerable. Therefore, numerous experimental results on MI require a more realistic theory taking into account a specific tensor form of the ac permeability and impedance.

The calculation of $\hat{\varsigma}$ is based on the solution of the Maxwell equations inside the conductor for the ac fields **e** and **h** together with the equation of motion for the magnetization vector $\mathbf{M}$. An analytical treatment is possible in a linear approximation with respect to the time-variable parameters $\mathbf{e}$, $\mathbf{h}$, and $\mathbf{m} = \mathbf{M} - \mathbf{M}_0$, where $\mathbf{M}_0$ is the static magnetization. Assuming a local linear relationship between $\mathbf{m}$ and $\mathbf{h}$: $\mathbf{m} = \hat{\chi}\,\mathbf{h}$, the problem is simplified to finding the solutions of the Maxwell equations with a given ac permeability matrix $\hat{\mu} = 1 + 4\pi\,\hat{\chi}$. In general, the anisotropy axis $\mathbf{n}_K$ has an angle $\psi$ with the wire axis (x-axis), as shown in Fig. 3. In the model approximation the wire is assumed to be in a single domain state with the static magnetization $\mathbf{M}_0$ directed in a helical way having an angle $\theta$ with the x-axis. The external dc magnetic field $H_{ex}$ is assumed to be parallel to the wire axis. The field applied in the perpendicular direction does not effect the magnetic configuration. Then, in the system of randomly oriented wires the magnetic properties need to be averaged over the field orientation.

The equilibrium direction of $\mathbf{M}_0$ is found by minimizing the magnetostatic energy $U$:[26]

$$\frac{\partial U}{\partial \theta} = 0, \tag{24}$$

$$U = -K\cos^2(\psi - \theta) - M_0 H_{ex} \cos(\theta),$$

where $K$ is the anisotropy constant and $\psi$ is the anisotropy angle. Equation (24) describes the rotational magnetization process. This single domain approach is known as the Stoner-Wohlfarth rotation model.[66] It is of interest to ascertain how the Stoner-Wohlfarth rotation model can be used for the analysis of the MI field behavior in real samples that always have a multidomain state. In the case of circumferential anisotropy in wires ($\psi = 90^0$), this initial multidomain state concerns the "bamboo-like" domain structure. When a dc longitudinal magnetic $H_{ex}$ field is applied to such a magnetic structure, the only magnetization rotation is the mechanism in charge of the reversal process. The centers of the domain walls remain immovable and only their widths increase with the simultaneous growth of the dominant magnetization along the direction $H_{ex}$ (due to its rotation). The corresponding dc magnetization curve $M_{0x}(H_{ex})$ (the projection of $\mathbf{M}_0$ on the wire axis) is anhysteretic with a linear dependence for $|H_{ex}| < H_K$ and approaches saturation for $|H_{ex}| > H_K$, where $H_K = 2K/M_0$ is the anisotropy field. The comparative analysis of the theoretical model and the measured impedance in wires with circumferential anisotropy and "bamboo-like" domain structure has been carried out in Ref. 50 over a wide frequency range. While the frequency does not exceed a few tens of megahertz ($< 50$ MHz), the ac domain susceptibility may introduce a considerable contribution to $\hat{\mu}$ within the field range $|H_{ex}| < H_K$, where the sample remains a multidomain structure. Obviously, this contribution can not be described in the frame of the rotation model, and therefore a disagreement between the measured and theoretical MI curves should be observed. However, in the field range $|H_{ex}| > H_K$, when the magnetization approaches saturation, the theoretical model will describe the experimental curve quite well. At higher frequencies the domain wall movement becomes strongly damped and contributes little to the ac magnetization. Thus, the ac matrix $\hat{\mu}$ can be determined by the magnetization rotation only, depending on the equilibrium direction of $\mathbf{M}_0$. As a consequence of this, the agreement between the theoretical and measured MI curves becomes better even for the field range $|H_{ex}| < H_K$. For the helical and axial anisotropies ($0 \le \psi < 90^0$) the magnetization process demonstrates a distinct hysteresis of $M_{0x}(H_{ex})$. The dc magnetization curve can be subdivided into two parts: reversible rotation and irreversible magnetization jumps, which are typically related to domain processes because the coercitivity field $H_c$ is much smaller than the anisotropy field $H_K$. Therefore, the rotation model can be used only as a certain extrapolation within the field interval $|H_{ex}| < H_c$.

As it has been mentioned above, the MI field characteristics at very high frequencies are almost provided by the magnetization rotation with the equilibrium angle $\theta(H_{ex})$. The type of MI field characteristics obtained strongly depends on the anisotropy angle $\psi$.[50-52] Besides, both the anisotropy constant $K$ and angle $\psi$ depend on the stress distribution inside the sample due to the inverse magnetostriction effect. In general, this stress distribution has a tensor character.[41-46] In glass-coated wires a large internal tensile stress $\sigma_0$ is introduced by the glass coating during the fabrication process. In conjunction with the negative magnetostriction ($\lambda < 0$) in Co-based wires the tensile stress $\sigma_0$ results in circumferential anisotropy and the "bamboo-like" domain structure (see Fig. 1) with the circular magnetization at $H_{ex} = 0$. Any torsion stress (internal or external) will deflect the equilibrium magnetization from the circular direction at $H_{ex} = 0$.

For potential applications concerned with the stress tunable composite materials only the external tensile stress is available to be transmitted to an individual wire inclusion through the composite matrix. In a tensile pre-stress composite matrix, an external compression will reduce only the initial tensile stress acting on the wire inclusions. To provide the tensile stress dependence of $\theta$, the equilibrium magnetization must be helical. In fact, the tensile stress induces the magnetic anisotropy in the direction of the stress in the case of a positive magnetostriction $\lambda > 0$ and in the transverse direction for $\lambda < 0$. The induced anisotropy defines the preferred direction of the static magnetization. At $H_{ex} = 0$, the tensile stress in both cases of $\lambda > 0$ and $\lambda < 0$ does not produce any change in the equilibrium magnetization reinforcing only the effective anisotropy field: the circumferential for $\lambda < 0$ and the longitudinal for $\lambda > 0$. Only in the case of a helical magnetization, when the equilibrium magnetization is deflected from the circumferential or longitudinal directions, the magnetization rotation due to the tensile stress becomes possible.

One possibility is to realize somehow a "frozen" torsion stress and corresponding helical anisotropy. For example, tension-annealed CoSiB amorphous wires of 30 μm in diameter (magnetostriction $\lambda \sim -10^{-6}$) posses a spontaneous helical anisotropy due to a residual stress distribution, and a relatively large anisotropy field.[67] Also, annealing under a torsion stress or current annealing in the presence of the axial magnetic field can induce a helical anisotropy.[52,68] In all the cases, the anisotropy angle is difficult to control. Therefore, for practical application the reliable method of inducing the helical anisotropy has to be developed. In another case, a helical magnetization can be achieved in wires with a circumferential anisotropy ($\lambda < 0$) by the combination of the external tensile stress $\sigma_{ex}$ and the fixed dc longitudinal magnetic field $H_{ex}$, which should be chosen about of the anisotropy field $H_K$. In this simplest case ($\psi = 90^0$), we can neglect the small internal torsion stresses in the wire in comparison of the initial tensile stress $\sigma_0$. Then, the anisotropy constant $K$ in Eq. (24) will be determined by the following relationship:

$$K = \frac{3}{2} |\lambda| (\sigma_{ex} + \sigma_0). \tag{25}$$

In glass-coated wires, a large internal tensile stress $\sigma_0$ is induced by the glass coating during the fabrication process. Turning back to Eq. (24), the external tensile stress $\sigma_{ex}$ reinforces the circumferential anisotropy, whereas $H_{ex}$ deflects the magnetization from the initial circumferential direction. These two factors allow the magnetization rotation due to $\sigma_{ex}$ at the fixed $H_{ex}$. A more elaborated model, including the initial torsion stresses, was considered in Ref. 24 in connection with stress tunable composites.

The total impedance tensor $\hat{\varsigma}$ has been calculated in Ref. 50 for the strong and weak skin-effects. Here we give only the two components $\varsigma_{xx}$ and $\varsigma_{x\varphi}$ ($\equiv \varsigma_{\varphi x}$) which are used in the generalized antenna equation (6):

a) strong skin-effect ($\delta_m / a << 1$)

$$\varsigma_{xx} = \frac{c(1-i)}{4\pi\sigma\delta}\left(\sqrt{\tilde{\mu}}\cos^2(\theta) + \sin^2(\theta) + \frac{\delta(1+i)}{4a}\right), \tag{26a}$$

$$\varsigma_{x\varphi} = \frac{c(1-i)}{4\pi\sigma\delta}\left(\sqrt{\tilde{\mu}} - 1\right)\sin(\theta)\cos(\theta),$$

b) weak skin-effect ($a/\delta << 1$) or ($a/\delta \sim 1$, $a/\delta_m \sim 1$)

$$\varsigma_{xx} = \frac{k_m c}{4\pi\sigma}\frac{J_0(k_m a)}{J_1(k_m a)} + \frac{1}{54}\left(\frac{a}{\delta}\right)^4 \frac{c\mu_3^2}{\pi\sigma a}, \tag{26b}$$

$$\varsigma_{x\varphi} = i\frac{a\omega}{3c}\mu_3 - \left(\frac{a}{\delta}\right)^4\left[\frac{\mu_1\mu_3}{60} + \frac{\mu_2\mu_3}{30}\right]\frac{c}{\pi\sigma a}.$$

Here $\tilde{\mu} = 1 + 4\pi\,\tilde{\chi}$ and $\tilde{\chi} = \chi_2 - 4\pi\,\chi_a^2/(1 + 4\pi\,\chi_1)$ are the effective permeability and susceptibility respectively, $\mu_1 = 1 + 4\pi\cos^2(\theta)\tilde{\chi}$, $\mu_2 = 1 + 4\pi\sin^2(\theta)\tilde{\chi}$, $\mu_3 = -4\pi\sin(\theta)\cos(\theta)\tilde{\chi}$, $k_m^2 = \mu_1\left(4\pi\, i\omega\sigma/c^2\right)$, $\delta_m = c/\sqrt{2\pi\sigma\omega\mu_1}$ is the magnetic skin-depth, $\delta = c/\sqrt{2\pi\sigma\omega}$ is the non-magnetic skin-depth, $\sigma$ is the wire conductivity, $J_{0,1}$ are the Bessel functions. The terms in the effective susceptibility $\tilde{\chi}$ are calculated from the following equations:

$\chi_1 = \omega_M(\omega_1 - i\tau\,\omega)/\Delta$,

$\chi_2 = \omega_M(\omega_2 - i\tau\,\omega)/\Delta$,

$\chi_a = \omega\,\omega_M/\Delta$,

$\Delta = (\omega_2 - i\tau\,\omega)(\omega_1 - i\tau\,\omega) - \omega^2$,

$\omega_1 = \gamma\left(H_{ex}\cos(\theta) + H_K\cos(2(\psi - \theta))\right)$,

$\omega_2 = \gamma\left(H_{ex}\cos(\theta) + H_K\cos^2(\psi - \theta)\right)$,

$\omega_M = \gamma M_0$, $H_K = 2K/M_0$,

where $\gamma$ is the gyromagnetic constant, $\tau$ is the spin-relaxation parameter, $H_K$ is the anisotropy field. Equations (26a) and (26b) demonstrate that the components of surface impedance tensor depend on both the ac susceptibility parameter $\tilde{\chi} = (\tilde{\mu} - 1)/4\pi$ and the static magnetization orientation angle $\theta$. At high frequencies the latter will give the main contribution to the field or tensile stress dependence of the impedance components, since $\tilde{\chi}$ looses its field sensitivity.

Figure 4 demonstrates the typical dispersion curves for the effective permeability parameter $\tilde{\mu} = 1 + 4\pi\,\tilde{\chi}$ that enters the impedance matrix in combination with the magnetization angle $\theta$. The following magnetic parameters have been chosen: anisotropy field $H_K = 2$ Oe, saturation magnetization $M_0 = 500$ G, gyromagnetic constant $\gamma = 2\times 10^7$ (rad/s)/Oe. In calculations, a small dispersion of the anisotropy angle $\psi$ with respect to $90^0$ should be introduced to model a real sample and avoid zero ferromagnetic resonance frequency at $H_{ex} = H_K$. The real part $\mathrm{Re}(\tilde{\mu}(\omega))$ approaches unity at the ferromagnetic resonance frequency ($\psi = 90^0$) $f_{FMR} = \gamma\sqrt{|H_{ex} - H_K| 4\pi M_0}\big/ 2\pi$ : $f_{FMR} = 365$ MHz at $H_{ex} = 0$ and $f_{FMR} = 725$ MHz at $H_{ex} = 10$ Oe. In the gigahertz range, $\mathrm{Re}(\tilde{\mu}(\omega))$ is negative being in magnitude in the range of 10–20, and $\mathrm{Im}(\tilde{\mu}(\omega))$ is in the range of 10–40. Both of them become insensitive to the external magnetic field (and the tensile stress), as shown in Fig. 4(c). In this case, the field and stress dependences of the impedance components are entirely related with that for the static magnetization orientation $\theta$. Then, if $\theta$ is a sensitive function of $H_{ex}$ or $\sigma_{ex}$, to insure high field sensitivity of the impedance, it is important only that the condition $|\tilde{\mu}(\omega)| >> 1$ is held (actually, $|\tilde{\mu}| \sim 10$ is enough). This conclusion clearly demonstrates that the condition of the ferromagnetic resonance is not required for the MI effect, contrary to the widely expressed belief.[69,70]

A large difference between $f_{FMR}$ and the frequency where the imaginary part reaches a maximum value is caused by the specific form of the effective susceptibility $\tilde{\chi}$ containing all the components of the susceptibility matrix $\hat{\chi}$ :[50]

$$\tilde{\chi} = \frac{\omega_M(\omega_2 - i\tau\omega) + 4\pi\omega_M^2}{(\omega_1 - i\tau\omega)(\omega_2 + 4\pi\omega_M - i\tau\omega) - \omega^2} \tag{27}$$

The expression for $f_{FMR}$ directly follows from Eq. (27): $f_{FMR} = \sqrt{\omega_1(\omega_2 + 4\pi\omega_M)}\big/ 2\pi$ . The dispersion curves, considered above, look very similar to a relaxation spectrum typical of polycrystalline multidomain ferrites. However, in our case, the “relaxation-like” dispersion is caused by a complicated form (27) of the effective susceptibility. Such kind of dispersion is always observed in experiments with bulk ferromagnetic conductive samples, where the skin-effect is important and the effective susceptibility is composed of the components of the internal matrix $\hat{\chi}$.[71]

Figure 5 demonstrates the typical field dependence of $|\varsigma_{xx}(H_{ex})|$ in the GHz range for a wire with circumferential anisotropy ($\psi = 90^0$). The wire has 10 μm diameter, conductivity $\sigma = 7.6\times10^{15}$ s$^{-1}$, anisotropy field $H_K = 2$ Oe, saturation magnetization $M_0 = 500$ G, and gyromagnetic constant $\gamma = 2\times10^7$ (rad/s)/Oe. The calculations have been carried out in a low frequency limit (see Eq. (26b)) since the magnetic skin-depth $\delta_m = c/\sqrt{2\pi\sigma\omega\mu_1}$ is of the order of the wire radius. As it will be shown below, a moderate skin effect is the basic requirement for obtaining the field/stress dependent effective permittivity. For this reason the wire diameter has to be chosen to be sufficiently small. At GHz frequencies the curve of $|\varsigma_{xx}(H_{ex})|$ flattens at $|H_{ex}| > H_K$ and even for large values of $|H_{ex}| >> H_K$ does not reach the saturation. In the GHz range $|\varsigma_{xx}(H_{ex})|$ has approximately a constant value for $|H_{ex}| \geq H_K$ reflecting the field dependence of $\cos^2(\theta)$ (see Eqs. (26a,b)) since the permeability parameter $\tilde{\mu}$ looses its field/stress sensitivity (as discussed above in Fig. 4(c)). Thus, at very high frequencies the impedance $|\varsigma_{xx}(H_{ex})| \sim \cos^2(\theta)$ exhibits a "valve-like" behavior, switching from one stable level to the other, following the dc magnetization. Such kind of the transformation of $|\varsigma_{xx}(H_{ex})|$ was previously reported in Ref. 72 where a rather confusing explanation of this effect was given. The first accurate measurements of the "valve-like" impedance were made in Ref. 73 for Co-rich glass-coated amorphous wires of $\sim 10$ μm in diameter with vanishing negative magnetostriction. The circular domain structure resulted in nearly linear non-hysteretic magnetization curves $M_{0x}(\theta) = \cos(\theta)$ providing the valve-like behavior $|\varsigma_{xx}(H_{ex})| \sim \cos^2(\theta)$. Figures 6(a) and 6(b) demonstrate similar experimental dependencies of the wire impedance $|Z(H_{ex})| \sim |\varsigma_{xx}(H_{ex})|$ measured in our present work in a CoMnSiB amorphous glass-coated wire[36-39] of $\sim 11-12$ μm in diameter with vanishing negative magnetostriction $\sim -2\times 10^{-7}$, the anisotropy field $H_K \sim 2.5$ Oe and the coercitivity field $H_c \sim 0.1-0.15$ Oe. To enable such high frequency impedance measurements (at 2.5 GHz and 3.9 GHz), the rf measurement cell requires special calibration as described in Ref. 24. The powerful calculation facilities of modern Analyzers can correct for the dispersion in the measurement track to create a virtually ideal passage characteristic. This allowed us to use a convenient design for the rf measurement cell. The field behaviors observed in Figs. 5 and 6 completely differ from that in the MHz range, where $|Z(H_{ex})| \sim |\varsigma_{ex}(H_{ex})|$ has two well-defined peaks at $H_{ex} \approx \pm H_K$ and decreases with $|H_{ex}| > H_K$ reaching the level of $\sim |Z(0)|$, as shown in Fig. 7.

The stress impedance in the GHz range was measured for the first time in Ref. 24 in the same CoMnSiB amorphous glass-coated wire. Figure 8 shows the plot of $|Z(\sigma_{ex})|$ vs. applied tensile stress $\sigma_{ex}$ for the frequency 2.5 GHz with $H_{ex}$ as a parameter. If no field is applied, the stress effect is almost not noticeable at GHz frequencies. In the presence of $H_{ex}$, the wire impedance shows large changes in response to the application of the tensile stress. The highest stress sensitivity is obtained for $H_{ex} = 3$ Oe that is about the same value as the anisotropy field for the unloaded wire. This impedance behavior vs. stress is in agreement with the previous discussion concerning Eqs. (24) and (25). Since a negative magnetostriction wire has a nearly circumferential anisotropy, the applied tensile stress alone will not cause the change in $\mathbf{M}_0$ direction, and as a result, will not produce noticeable changes in the impedance at high frequencies. Applying the field $H_{ex}$ of the order of $H_K$ saturates the wire in the axial direction. The tensile stress which enlarges the circumferential anisotropy in the case of negative magnetostriction acts in opposite way and rotates the magnetization back to the circular direction. The highest stress sensitivity is obtained for $H_{ex} \sim H_K$, which is sufficient to saturate the wire. Further increase in $H_{ex}$ is unnecessary since a larger stress will be required to return the magnetization back. The theoretical plots showing $|\varsigma_{xx}(\sigma_{ex})| \sim |Z(\sigma_{ex})|$ for frequency of 2.5 GHz with $H_{ex}$ as a parameter are shown in Fig. 9, demonstrating very good agreement with the experimental results. The initial tensile stress and magnetostriction constant were chosen to be $\sigma_0 = 200$ MPa and $\lambda \sim -2\times10^{-7}$ respectively, to provide a value of $H_K \sim 2.5$ Oe. Within calculations, the simplest relationship (25) can be used, since the initial tensile stress $\sigma_0$ (due to the glass coating) prevails over the possible small torsion stresses inside the wire.

The direct measurement of the off-diagonal component $\varsigma_{x\varphi}(H_{ex},\sigma_{ex})$ of the impedance tensor in the GHz range is an intractable problem, although in the MHz range the field dependencies all of the tensor components have been successfully measured.[51,52] Therefore, the measurement of the effective properties of the whole composite is the only possible method that is available to investigate the effects related with the off-diagonal tensor components.

## V. Field and stress dependencies of the effective permittivity

The field or stress dependences of the effective permittivity $\varepsilon_{eff}$ of the composite is caused by the field/stress dependence of the surface impedance matrix $\hat{\varsigma}(H_{ex},\sigma_{ex})$, which determines the losses inside the inclusions (see Eq. (2)). These internal losses characterize the quality factor of the entire composite system and the type of dispersion of $\varepsilon_{eff}(\omega)$. It is quite natural to expect that the field/stress dependence of $\varepsilon_{eff}(\omega)$ becomes most sensitive in the vicinity of the antenna resonance where any small variations in the inclusion parameters cause a strong change of the current distribution and the inclusion dipole moment. This results in a remarkable transformation of the dispersion curve $\varepsilon_{eff}(\omega)$ under the external dc magnetic field or tensile stress.

The resonance frequency range is determined by the wire length $l$ and the matrix permittivity $\varepsilon$. Practically, it is not desirable to construct composite materials with inclusions having a length larger than 1–2 cm. In this case, the first resonance frequency in air $f_{res} = c/2l$ would be in the range of tens gigahertz. However, for such high frequencies the magnetic properties of soft ferromagnetic wires tail off completely and $\tilde{\mu}$ in Eqs. (26a,b) tends to be unity. Without increasing the wire length, the operating frequencies can be lowered in $\sqrt{\varepsilon}$ times by using a dielectric matrix with higher permittivity $\varepsilon >> 1$: $f_{res} = c/2l\sqrt{\varepsilon}$. Some polymers and commercial epoxies can be used as a dielectric matrix. A fine-dispersion filler (powder) containing particles with a large polarisability can be used to further increase $\varepsilon$, for example, the powder of ceramic microparticles. Another method of increasing $\varepsilon$ uses fine-disperse metal powder.[9] Both methods allow the matrix permittivity $\varepsilon$ to be made very large with a small tangent of losses.

The condition of a moderate skin-effect ($a/\delta_m \sim 1$) is proving to be important to realize a high sensitivity of $\varepsilon_{eff}(\omega)$ to the external factors ($H_{ex}$ and $\sigma_{ex}$). If the magnetic skin-depth is much smaller than the wire radius ($\delta_m / a << 1$), the normalized wave number $\tilde{k}$ (see Eq. (14)) differs little from the wave number $k$ of the free space. Substituting the high frequency impedance (26a) into Eq. (14) gives:

$$\tilde{k} \sim \frac{\omega}{c}\sqrt{\varepsilon\mu}\left(1 + \frac{(1+i)}{2\mu\,\ln(l/a)}\frac{\delta}{a}\sqrt{\tilde{\mu}}\cos^2(\theta)\right)^{1/2}. \tag{28}$$

From Eq. (28) it immediately follows that if $\delta / a << 1$ (and $\delta_m / a << 1$) the wave number becomes $\tilde{k} \sim \omega\sqrt{\varepsilon\mu}/c$, hence it follows that an essential field/stress dependence of $\varepsilon_{eff}(\omega)$ can be reached for the case when the non-magnetic skin depth is of the order of wire radius.

At a very low inclusion concentration ($p << p_c \sim 2a/l$) the effective permittivity $\varepsilon_{eff}(\omega)$ can be represented by Eq. (3) as the dipole sum with the polarisability $< \alpha >$ averaged over the inclusion orientations. The polarisability $\alpha$ has to be calculated from Eq. (1) for $j(x)$ evaluated from Eq. (6). Alternatively, the first approximation $j_1(x)$, which is evaluated by means of the iteration procedure developed in Refs 23 and 54, can be used to take into account all the losses in the system. In principle, the asymptotic expression for $j(x)$ (obtained within this iteration procedure) can be written in the form of the non-local Ohm's law: $j(x) = (\hat{\sigma} * \bar{e}_{0x})$ – is the convolution with the external electric field $\bar{e}_{0x}(x)$ ($\bar{h}_{0x} \equiv 0$) and $\hat{\sigma}(x,s)$ is the kernel of the conductivity integral operator acting within the volume occupied by a wire (thin tube). The corresponding kernel for the permittivity integral operator takes the form $\hat{\varepsilon}(x,s) = 4\pi\, i\hat{\sigma}(x,s)/\omega$. Averaging this non-local permittivity operator over the whole sample, we obtain the non-local operator of the effective permittivity – even for the single particle approximation! In our present work we do not elaborate upon this subject, since all observed effects can be understood in the frame of a simple model using Eq. (3). Nevertheless, the effect of spatial dispersion may become significant at optical frequencies in composites filled with the nano-wire inclusions.

For a composite layer filled with the non-chiral wire inclusions ($\varsigma_{x\varphi} \equiv 0$ in Eq. (6)) the total permittivity tensor $\hat{\varepsilon}_{eff}$, characterizing its effective dipole response, is of the following form ($p << p_c$):

$$\hat{\varepsilon}_{eff} = \begin{pmatrix} \varepsilon_{11} & 0 & 0 \\ 0 & \varepsilon_{22} & 0 \\ 0 & 0 & \varepsilon_{33} \end{pmatrix}, \qquad (29)$$

where $\varepsilon_{11}$ and $\varepsilon_{22}$ are the effective permittivities in the plane of sample $(x, y)$ for mutually perpendicular directions. For a low inclusion concentration ($p << p_c$), the effective permittivity $\varepsilon_{33}$ in z-direction is equal to the matrix permittivity ($\varepsilon_{33} \approx \varepsilon$) since ac electric field perpendicular to the wire inclusions (and the sample surface) does not cause a significant dipole polarization. For a composite layer with random orientation of the wire inclusions shown in Fig. 2, $\varepsilon_{11} = \varepsilon_{22} = \varepsilon_{eff}$ and the averaging $< \alpha >$ over the inclusion orientations in Eq. (3) gives a coefficient $1/2$ ($< \alpha >= \alpha / 2$). To the contrary, in the composite sample with the ordered wire inclusions (in x-direction, for example), $\varepsilon_{11} = \varepsilon_{eff}$ with $< \alpha >= \alpha$, whereas $\varepsilon_{22} = \varepsilon_{33} = \varepsilon$.

Figure 10 demonstrates the theoretical dispersion curves of the effective permittivity $\varepsilon_{eff}(\omega)$ in the GHz range for the composite structure shown in Fig. 2 ($\varepsilon_{11} = \varepsilon_{22} = \varepsilon_{eff}$, $\varepsilon_{33} = \varepsilon$, $< \alpha >= \alpha / 2$). The dc external magnetic field $H_{ex}$, used as a parameter, was applied along the sample surface at any azimuthal direction. The matrix permittivity was $\varepsilon = 16$ and volume concentration $p = 0.001\%$. This concentration is considerably smaller than the percolation threshold for such composites. For a sufficiently large external dc field $|H_{ex}| >> H_K$ its azimuthal orientation $\phi$ in the sample plane is of no importance because of the "valve-like" behavior of MI (see Figs. 5 and 6). In fact, with $|H_{ex}| >> H_K$ the magneto-impedance $|\varsigma_{xx}|$ will be switched to the same state for the majority of wires since the condition $|H_{ex}\cos(\phi)| > H_K$ will hold true for most wire orientations, where $|H_{ex}\cos(\phi)|$ is the projection of $H_{ex}$ on a wire. Therefore, the "valve-like" behavior of the MI is very important to provide the homogeneous field-tunable properties in composites with random orientation of the wire inclusions. The field-dependence effect in Fig. 10 shows up in changing character of the dispersion curves. In the absence of $H_{ex}$ the dispersion curve is of a resonance type. Applying a sufficiently large $|H_{ex}| > H_K$, the impedance $\varsigma_{xx}(H_{ex})$ is increased and, as a consequence, the internal losses in the inclusion, which results in the dispersion of a relaxation type. In the presence of $H_{ex}$ the resonance frequency also slightly shifts towards higher frequencies. In Ref. 8 the transformation of the dispersion $\varepsilon_{eff}(\omega)$ from a resonant type to relaxation one was associated with a different wire conductivity, which defines the resistive loss. In our case it is provided through the field dependent impedance $\varsigma_{xx}(H_{ex})$ instead of conductivity. At a wire concentration larger than a certain value, the real part of $\varepsilon_{eff}(\omega)$ may become negative in the vicinity of the resonance.[9,23] Therefore, applying $H_{ex}$, it is possible change gradually $\mathrm{Re}(\varepsilon_{eff}(\omega))$ from negative to positive values.[23]

The composite structure allowing stress-tunable properties should be composed of ordered wire inclusions, since the critical dc bias field $H_{ex}$ providing the sensitive stress dependence of MI must have the same value $H_{ex} \sim H_K$ for all wires with a circumferential anisotropy (see Figs. 8 and 9). Figure 11 shows the theoretical dispersion curves for the effective permittivity $\varepsilon_{eff}(\omega)$ ($\varepsilon_{11} = \varepsilon_{eff}$, $\varepsilon_{22} = \varepsilon_{33} = \varepsilon$, $<\alpha> = \alpha$) with $H_{ex}$ and $\sigma_{ex}$ as external parameters. The matrix permittivity was $\varepsilon = 16$ and the volume concentration $p = 0.0005\%$. With $H_{ex} \sim H_K$ applied along the wire inclusions, $\varepsilon_{eff}(\omega)$ demonstrates the strong dependence on the external tensile stress $\sigma_{ex}$. This characteristic feature of the tensile stress dependence of $\varepsilon_{eff}(\omega)$ is a direct consequence of the corresponding behavior of $\varsigma_{xx}(\sigma_{ex})$ demonstrated in Figs. 8 and 9.

For a composite layer filled with wire inclusions with a helical anisotropy ($\varsigma_{x\varphi} \neq 0$ in Eq. (6)), the bulk electrical polarization $\mathbf{P}$ may exhibit a chiral property,[53] when it becomes proportional to both the electrical $\mathbf{e}_0$ and magnetic $\mathbf{h}_0$ excitation fields:[3,4]

$$\mathbf{P} = (\vartheta \mathbf{e}_0 + \vartheta_{eff} \mathbf{e}_{0\|}) + \beta_{eff} \mathbf{h}_{0\|}, \tag{30}$$

where $\vartheta$ is the electrical susceptibility of the dielectric matrix, $\vartheta_{eff}$ and $\beta_{eff}$ are the effective bulk susceptibility due to the wire dipole response induced by the field projections $\mathbf{e}_{0\|}$ and $\mathbf{h}_{0\|}$ in the plane of the composite layer. For the electric intensity vector $\mathbf{d}$ inside the composite layer we obtain the following representation:

$$\mathbf{d} = \mathbf{e}_0 + 4\pi\, \mathbf{P} = (1 + 4\pi\vartheta)\mathbf{e}_0 + 4\pi\vartheta_{eff} \mathbf{e}_{0\|} + 4\pi\, \beta_{eff} \mathbf{h}_{0\|} \tag{31}$$

or

$$\mathbf{d} = \varepsilon\, \mathbf{e}_0 + 4\pi\vartheta_{eff} \mathbf{e}_{0\|} + 4\pi\, \beta_{eff} \mathbf{h}_{0\|},$$

where $\varepsilon = 1 + 4\pi\, \vartheta$ is the matrix permittivity. If $\mathbf{e}_0$ and $\mathbf{h}_0$ are polarized in the plane of the composite slab ($\mathbf{e}_0 \equiv \mathbf{e}_{0\|}$ and $\mathbf{h}_0 \equiv \mathbf{h}_{0\|}$) we can write:

$$\mathbf{d}_{\|} = \varepsilon_{eff} \mathbf{e}_0 + 4\pi\, \beta_{eff} \mathbf{h}_0, \tag{32}$$

where $\varepsilon_{eff} = \varepsilon + 4\pi\, \vartheta_{eff}$ is the effective permittivity of composite. Since $|\varsigma_{x\varphi}| << 1$ (and hence $|\beta_{eff}| << 1$), the electrical excitation ($\varepsilon_{eff}$) will prevail over the magnetic one ($4\pi\, \beta_{eff}$) for most polarizations of the incident electromagnetic wave. Nevertheless, there are certain polarizations and composite microstructures when the magnetic excitation becomes important. For example, the electrical excitation is absent in a composite layer, when the

electromagnetic wave propagates along the composite surface and the electrical field is perpendicular to it, as shown in Fig. 12(a). Another configuration providing only the magnetic excitation is obtained for the composite where all wires are ordered in the same direction and the electrical field in the incident wave is perpendicular to this direction, as shown in Fig. 12(b).

As a conclusion, it should be noted that the chiral properties of a composite filled with the non-magnetic conductive wire inclusions become available owing to the wire spatial configurations (2D or 3D), such as spirals or omega-particles.[3,4,13,14] For ferromagnetic wires these spatial degrees of freedom can be replaced with an additional degree provided by the internal gyromagnetic properties of wires. In turn, this enables the one-dimensional chiral particles, such as the straight wires discussed above.

## VI. Potential applications

A number of applications can be proposed where the tunable properties of the effective permittivity and its selective absorption are of principal importance. The field-tunable composites may form selective microwave coatings with field-dependent reflection/transmission coefficients. The energy absorption in such kinds of composites is rather strong, therefore it has to be sufficiently thin to be used as a wave passage. Another promising application suggests the employment of these composites as an internal cover in partially filled waveguides or layered waveguides with a “dielectric/composite” structure. The waveguide with an internal composite cover will prove to operate similar to the waveguide containing the absorption ferromagnetic layers,[74] for example, a thin-film “dielectric/Fe” structure considered in Refs. 75 and 76. In both systems there is an anisotropic field-dependent layer with resonance properties and large energy losses, but in our case the field-dependent layer is made of a composite material with $\hat{\varepsilon}_{eff}$ . The operating frequency range is determined by the antenna resonance. Such a waveguide system, by analogy to that already designed for the waveguides with ferromagnets, can be used for tunable filters and phase shifters operating up to tens of gigahertz.

The use of microwaves for the condition assessment of structural elements is becoming established as a non-destructive evaluation method in civil engineering, especially for detection of invisible structural damages or defects. This technology is based on a reconstruction of dielectric profile (image) of a structure irradiated with microwaves, through scattering measurement controlled by software. The controllable changes of the scattering response, and hence the damage detection are possible only for adequate contrast variations in the material structure such as wide cracks or voids. Nevertheless, an interpretation of the microwave profile remains a very serious problem. Moreover, since pre-damage stress does not result in a felt variation in the dielectric constant of usual structural materials, the excess stress and material fatigue become unpredictable.

Since the mechanical stress is static in nature, an additional mediate physical process is required to visualize it. In our present work we have considered a new composite medium, which may visualize mechanical stress in the GHz range at any stage: before and after damage. The main feature of the proposed stress-tunable composite is its microwave permittivity, which depends on tensile stress. This kind of composite material can be characterized as a “sensing medium” and opens up new possibilities for remote monitoring of stress with the use of microwave transceiving techniques. The composite material can be made as a bulk medium or as a thin cover to image the mechanical stress distribution inside construction or on its surface. A possible design of monitoring system could use a network of stress-sensitive composite “blocks” embedded into the structure. The monitoring could be organized by means of radar scanning over the whole structure, when a microwave beam subsequently interrogates all the embedded sensing blocks over some frequency range which includes the resonance frequency. The reflection resonance spectrum from each block would be proportional to the acting stress. The total stress distribution in the structure can be restored using the contrast of the microwave absorption picture.

The free-space microwave technique operating in the far-field region has been employed for detecting voids and debonding between the polymer jacket and the concrete column.[77,78] This jacketing technology using fiber reinforced polymer (FRP) composites is applied for retrofit of reinforced concrete columns. The microwave imaging technology is based on the reflection analysis of a continuous electromagnetic wave sent toward and reflected from a layered FRP-concrete medium. Poor bonding conditions including voids and debonding will generate air gaps which produce additional reflections of the electromagnetic wave. No doubt, this method can be adopted for the application of stress-tunable composites considered in our present work.

Finally, it is necessary to distinguish the two main methods in non-destructive stress monitoring. The first one is based on the sensor network embedded into the structure where the distributed sensors characterize a local stress state. This method requires the connections with the readout electronics. The second method assumes the creation of a sensing medium, which can be characterized by some effective material parameters, such as the effective conductivity, permittivity or permeability. One of the significant advantages of such sensing media is the ability to use them in remote monitoring, for instance, by means of microwave scanning. Furthermore, this scanning method is not restricted with the difficulties of sending data through communication channels.

## ACKNOWLEDGMENTS:

The authors would like to acknowledge the contributions of the following participants of this work:

**Mr Nicholas Fry and Dr Serghey Sandacci**, for experimental work and manuscript preparation.

**Prof David Jiles** (The University of Iowa, USA), for his ideas relating to non-destructive testing of civil constructions presented at The University of Plymouth in Summer 2003.

**References:**

[1]A. K. Sarychev and V. M. Shalaev, Physics Reports **335**, 275-371 (2000).

[2]A. P. Vinogradov, *"Electrodynamics of Composite Materials"* (in Russian) (URSS, Russia, 2001).

[3]A. N. Serdyukov, I. V. Semchenko, S. A. Tretyakov, and A. Sihvola, *"Electromagnetics of Bi-anisotropic Materials: Theory and Applications"* (Amsterdam: Gordon and Breach Science Publishers, 2001).

[4]S. A. Tretyakov, *"Analytical Modeling in Applied Electromagnetics"* (Norwood, MA: Artech House, 2003).

[5]A. K. Sarychev, R. C. McPhedran, and V. M. Shalaev, Phys. Rev. **B 62**, 8531 (2000).

[6]R. Hilfer, Phys. Rev. **B 44**, 60 (1991).

[7]R. Hilfer, B. Nost, E. Haslund, Th. Kautzsch, B. Virgin, and B. D. Hansen, Physica **A 207**, 19 (1994).

[8]A. N. Lagarkov and A. K. Sarychev, Phys. Rev. **B 53**, 6318 (1996).

[9]A. N. Lagarkov, S. M. Matytsin, K. N. Rozanov, and A. K. Sarychev, J. Appl. Phys. **84**, 3806 (1998).

[10]D. P. Makhnovskiy, L. V. Panina, D. J. Mapps, and A. K. Sarychev, Phys. Rev. **B 64**, 134205 (2001).

[11]H. C. van de Hulst, *"Light Scattering by Small Particles"* (Dover, New York, 1981).

[12]C. F. Bohren and D. R. Huffman, *"Absorption and Scattering of Light by Small Particles"* (Wiley, New York, 1983).

[13]A. N. Lagarkov, V. N. Semenenko, V. A. Chistyaev, D. E. Ryabov, S. A. Tretyakov, and C. R. Simovski, Electromagnetics **17**, 213 (1997).

[14]I. V. Semchenko, S. A. Khakhomov, S. A. Tretyakov, and A. H. Sihvola, Electromagnetics **21**, 401 (2001).

[15]S. A. Baranov, Tech. Phys. Letters **24**, 549 (1998).

[16]O. Reynet, A.-L. Adent, S. Deprot, O. Acher, and M. Latrach, Phys. Rev. **B 66**, 094412 (2002).

[17]Most of the publications on this subject are collected at the web site of Massachusetts Institute of Technology: http://ceta-p5.mit.edu/metamaterials/papers/papers.html. Also we recommend the on-line network organized by Dr Enrico Prati, where the necessary information on metamaterials can be found: www.metamaterials.net .

[18]L. V. Panina and K. Mohri, Appl. Phys. Letters **65**, 1189 (1994); L. V. Panina, K. Mohri, K. Bushida, and M. Noda, J. Appl. Phys. **76**, 6198 (1994).

[19]K. Mohri, L. V. Panina, T. Uchiyama, K. Bushida, and M. Noda, IEEE Trans. Magn. **31**, 1266 (1995).

[20]N. Kawajiri, M. Nakabayashi, C. M. Cai, K. Mohri, and T. Uchiyama, IEEE Trans. Magn. **35**, 3667 (1999).

[21]K. Mohri, T. Uchiyama, L. P. Shen, C. M. Cai, and L. V. Panina, Sensors and Actuators **A91**, 85 (2001).

[22]K. Mohri, T. Uchiyama, L. P. Shen, C. M. Cai, L. V. Panina, Y. Honkura, and M. Yamamoto, IEEE Trans. Magn. **38**, 3063 (2002).

[23]D. P. Makhnovskiy and L. V. Panina, J. Appl. Phys. **93**, 4120 (2003).

[24]L. V. Panina, S. I. Sandacci, and D. P. Makhnovskiy, J. Appl. Phys. **97**, 013701 (2005).

[25]W. J. Biter, S. M. Hess, S. Oh, A. Geshury, and D. Heider, *The Proceeding of the Midwest Materials and Process Conference*, Covina, CA, September 2000; W. J. Biter, S. M. Hess, and S. Oh, *The Proceeding of the 33-rd International SAMPE Technical Conference*, Covina, CA, November 2001. Both publications are available at the web site of Sensortex Inc: www.sensortex.com

[26]S. Chikazumi, "*Physics of Magnetism*" (John Wily and Sons Inc., New York, London, Sydney 1964).

[27]D. Jiles, *"Introduction to Magnetism and Magnetic Materials"* (Chapman and Hall, London, UK, 1991).

[28]Y. Okuhara, B.-K. Jang, H. Matsubara, and M. Sugita, Proc. SPIE **5057**, 54 *"Smart Structures and Materials"* (2003).

[29]H. Chiriac and T. A. Ovari, Progress in Mater. Science **40**, 333 (1996).

[30]H. Chiriac, T.-A. Ovari, C.S. Marinescu, and V. Nagacevschi, IEEE Trans. Magn. **32**, 4755 (1996).

[31]H. Chiriac, T.-A. Ovari, Gh. Pop, and F. Barariu, J. Appl. Phys. **81**, 5817 (1997).

[32]H. Chiriac, T.-A. Ovari, Gh. Pop, and F. Barariu, IEEE Trans. Magn. **33**, 782 (1997).

[33]A. P. Zhukov, J. M. Blanco, J. Gonzalez, M. J. Garcia Prieto, E. Pina, and M. Vazquez, J. Appl. Phys. **87**, 1402 (2000).

[34]A. P. Zhukov, J. Gonzalez, J. M. Blanco, M. Vazquez, and V. S. Larin, J. Mater. Res. **15**, 2107 (2000).

[35]V. Zhukova, A. F. Cobeno, E. Pina, A. P. Zhukov, J. M. Blanco, L. Dominguez, V. S. Larin, and J. Gonzalez, J. Magn. Magn. Mater. **215-216**, 322 (2000).

[36]H. Chiriac, Gh. Pop, T.-A. Ovari, F. Barariu, M. Vazquez, and A. P. Zhukov, IEEE Trans. Magn. **33**, 3346 (1997).

[37]A. P. Zhukov, J. Gonzalez, J. M. Blanco, M. J. Prieto, E. Pina, and M. Vazquez, J. Appl. Phys. **87**, 1402 (2000).

[38]F. Cobeno, J. M. Blanco, A. P. Zhukov, and J. Gonzalez, J. Magn. Magn. Mater. **249/1-2**, 396 (2002).

[39]F. Cobeno, A. P. Zhukov, J. M. Blanco, and J. Gonzalez, J. Magn. Magn. Mater. **234**, L359-L365 (2001).

[40]V. S. Larin, A. V. Torcunov, A. P. Zhukov, J. Gonzalez, M. Vazquez, L. V. Panina, J. Magn. Magn. Mater. **249/1-2**, 39 (2002). See also: E. Ya. Badinter, *"Application of cast microwire in tool-making industry"*, A review article at the web site of "ELIRI Institute", Republic of Moldova. Download the Word File at the foot of the web page http://eliri.md/eng/about.htm

[41]H. Chiriac, T.-A. Ovari, and Gh. Pop, Phys. Rev. **B 52**, 10104 (1995).

[42]H. Chiriac, T.-A. Ovari, Gh. Pop, and F. Barariu, J. Magn. Magn. Mater. **160**, 237 (1996).

[43]E. Hristoforou, H. Chiriac, M. Neagu, I. Darie, and T.-A. Ovari, IEEE Trans. Magn. **32**, 4953 (1996).

[44]J. M. Blanco, A. P. Zhukov, and J. Gonzalez, J. Appl. Phys. **87**, 4813 (2000).

[45]J. M. Blanco, A. P. Zhukov, A. P. Chen, A. F. Cobeno, A. Chizhik, and J. Gonzalez, J. Phys. D: Appl. Phys. **34**, L31 (2001).

[46]V. Zhukova, J. M. Blanco, A. P. Zhukov, and J. Gonzalez, J. Phys. D: Appl. Phys. **34**, L113 (2001).

[47]M. Born and E. Wolf, *"Principles of Optics"* (Fourth edition, Pergamon Press, 1968).

[48]R. King and G. Smith, "*Antennas in Matter. Fundamentals, Theory and Applications*" (The MIT Press, Cambridge, Massachusetts, and London, England, 1981).

[49]L. D. Landau and E. M. Lifshitz, "*Electrodynamics of Continuous Media*" (Pergamon Press, 1975).

[50]D. P. Makhnovskiy, L. V. Panina, and D. J. Mapps, Phys. Rev. **B 63**, 144424 (2001).

[51]D. P. Makhnovskiy, L. V. Panina, and D. J. Mapps, J. Appl. Phys. **87**, 4804 (2000).

[52]V. A. Zhukova, A. B. Chizhik, J. Gonzalez, D. P. Makhnovskiy, L. V. Panina, D. J. Mapps, and A. P. Zhukov, J. Magn. Magn. Mater. **249/1-2**, 324 (2002).

[53]L. V. Panina, D. P. Makhnovskiy, and K. Mohri, J. Magn. Magn. Mater. **272-276/2**, 1452 (2004).

[54]L. V. Panina, A. N. Grigorenko, and D. P. Makhnovskiy, Phys. Rev. **B 66**, 155411 (2002).

[55]P. A. Belov, C. R. Simovski, and S. A. Tretyakov, Phys. Rev. **E 66**, 036610 (2002).

[56]P. A. Belov and S. A. Tretyakov, J. Electromag. Waves and Appl. **16**, 129 (2002).

[57]P. A. Belov, S. A. Tretyakov, and A. J. Viitanen, J. Electromag. Waves and Appl. **16**, 1153 (2002).

[58]S. A. Tretyakov and A. J. Viitanen, J. of the Optic. Soc. of America **A 17**, 1791 (2000).

[59]J. M. Ziman, *"Principles of the theory of solids"* (2d edition, Cambridge at the University Press, 1972).

[60]X. Xia, P. Yang, Y. Sun, Y. Wu, B. Mayers, B. Gates, Y. Yin, F. Kim, and H. Yan, Adv. Mater. **15**, 353 (2003).

[61]J. Aizpurua, P. Hanarp, D. S. Sutherland, M. Kall, G. W. Bryant, and F. J. Garcia de Abajo, Phys. Rev. Letters **90**, 057401 (2003).

[62]A. P. Vinogradov, D. P. Makhnovskiy, and K. N. Rozanov, J. of Commun. Techn. and Electronics **44**, 317 (1999).

[63]S. M. Matitsine, K. M. Hock, L. Liu, Y. B. Gan, A. N. Lagarkov, and K. N. Rozanov, J. Appl. Phys. **94**, 1146 (2003).

[64]D. P. Makhnovskiy, A. S. Antonov, A. N. Lagarkov, and L. V. Panina, J. Appl. Phys. **84**, 5698 (1998).

[65]D. P. Makhnovskiy, A. N. Lagarkov, L. V. Panina, and K. Mohri, Sensor and Actuators **A81**, 106 (2000).

[66]E. C. Stoner and E. P. Wohlfarth, Phil. Trans. Roy. Soc. **A240**, 599 (1948).

[67]L. V. Panina, H. Katoh, and K. Mohri, IEEE Trans. Magn. **29**, 2524 (1993).

[68]J. M. Blanco, A. P. Zhukov, and J. Gonzalez, J. Phys. D: Appl. Phys. **32**, 3140 (1999).

[69]D. Menard, M. Britel, P. Ciureanu, A. Yelon, V. P. Paramonov, A. S. Antonov, P. Rudkowski, and J. O. Strem-Olsen, J. Appl. Phys. **81**, 4032 (1997).

[70]M. R. Britel, D. Menard, L. G. Melo, P. Ciureanu, A. Yelon, R. W. Cochrane, M. Rouabhi, and B. Cornut, Appl. Phys. Lett. **77**, 2737 (2000).

[71]O. Acher, P.-M. Jacquart, and C. Boscher, IEEE Trans. Magn. **30**, 4542 (1994).

[72]M. Dominguez, J. M. Garcia-Beneytez, M. Vazquez, S. E. Lofland, and S. M. Bhagat, J. Magn. Magn. Mater. **249/1-2**, 117 (2002).

[73]S. I. Sandacci, D. P. Makhnovskiy, and L. V. Panina, J. Magn. Magn. Mater. **272-276/3**, 1855 (2004).

[74]A. G. Gurevich, "*Ferrites at Microwave Frequencies*" (New York: Consultants Bureau, 1963).

[75]E. Schloemann, R. Tutison, J. Weissman, H. J. Van Hook, and T. Vatimos, J. Appl. Phys. **63**, 3140 (1988).

[76]R. J. Astalos and R. E. Camley, J. Appl. Phys. **83**, 3744 (1998).

[77]M. Q. Feng, C. Liu, X. He, M. Shinozuka, J. Eng. Mech. (ASCE) **126**, 725 (1999).

[78]M. Q. Feng, Y. J. Kim, and F. De Flaviis, J. Eng. Mech. (ASCE) **128**, 172 (2001).

# Figures:

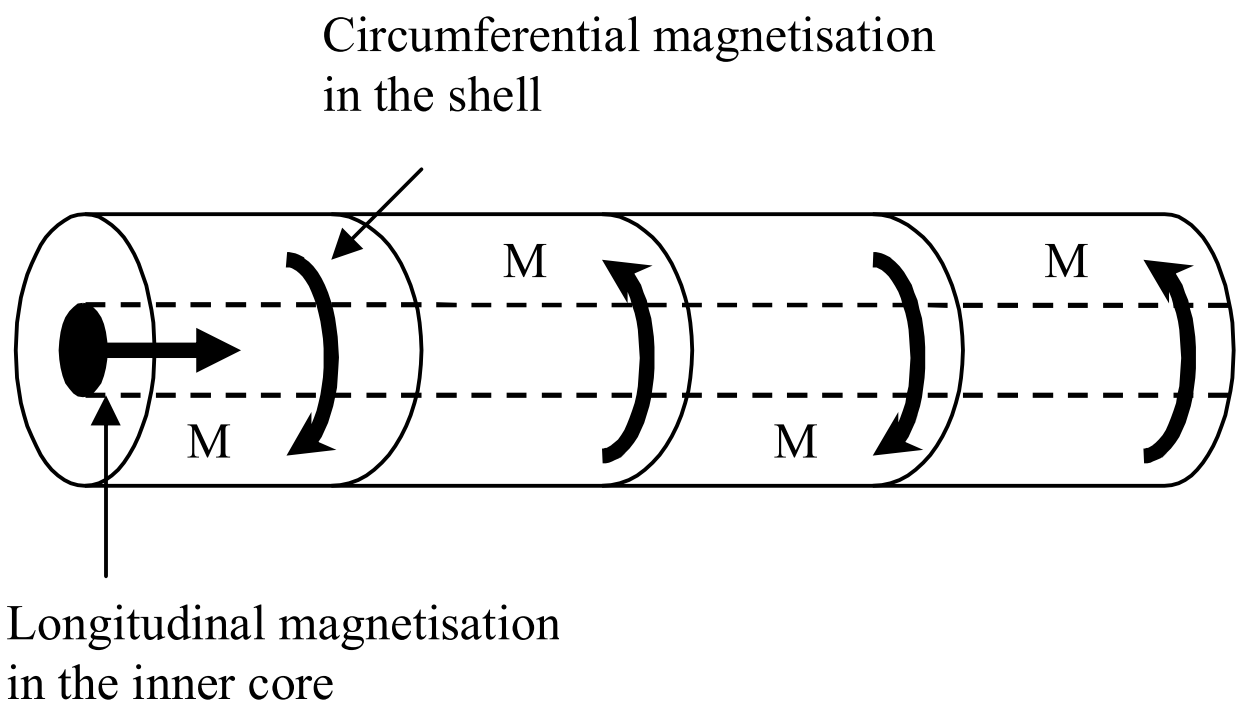

Fig. 1. Domain structure of a wire with negative magnetostriction ($\lambda < 0$). The wire sample with circumferential anisotropy ($\lambda < 0$) is divided into a "bamboo-like" domain structure, where adjacent domains have opposite directions of magnetization.

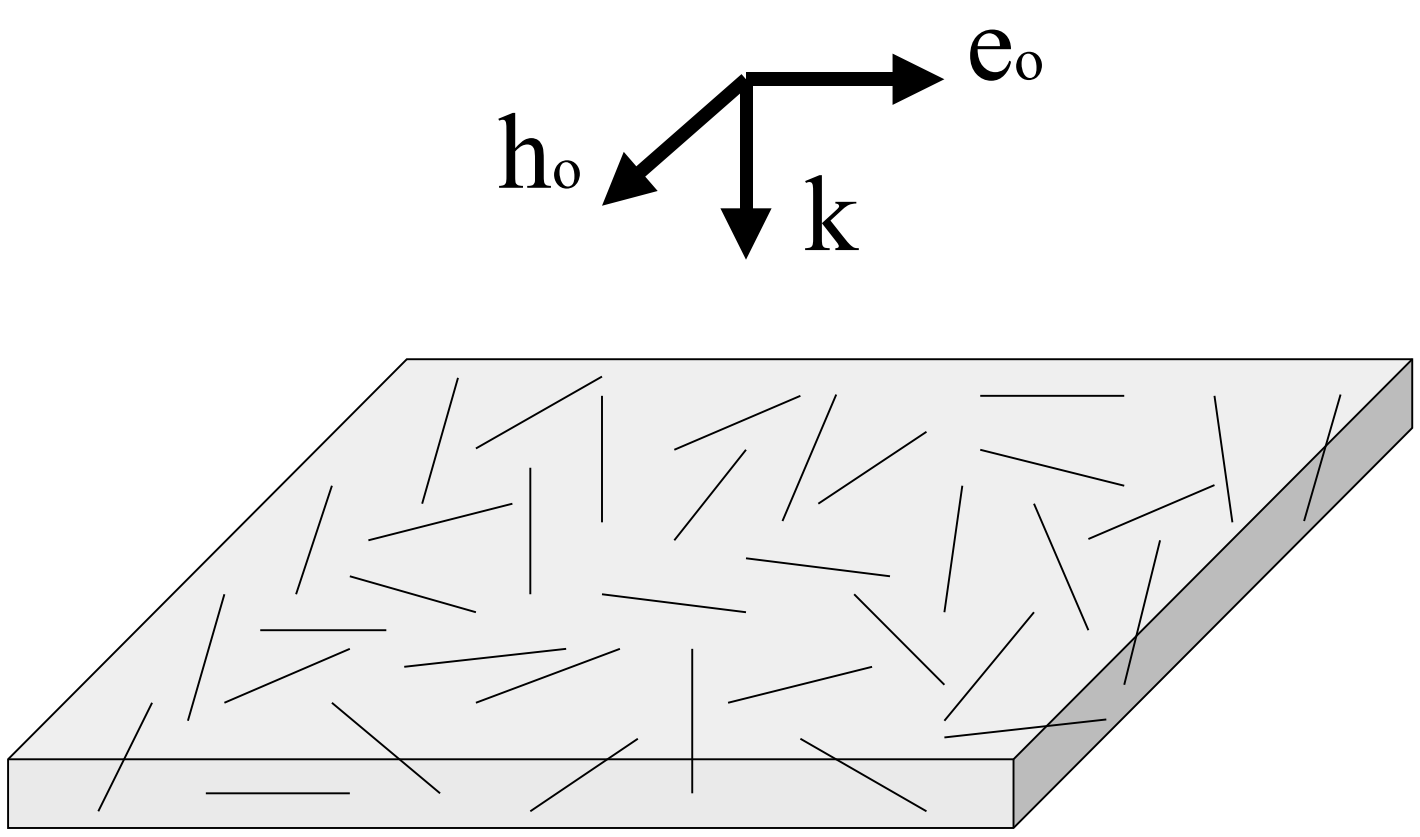

Fig. 2. Microstructure of a composite layer with a random orientation of wire inclusions in the plane of the sample. The composite is composed of the wire inclusions embedded into a dielectric matrix. The layer thickness $h$ is much smaller than the wire length $l$.

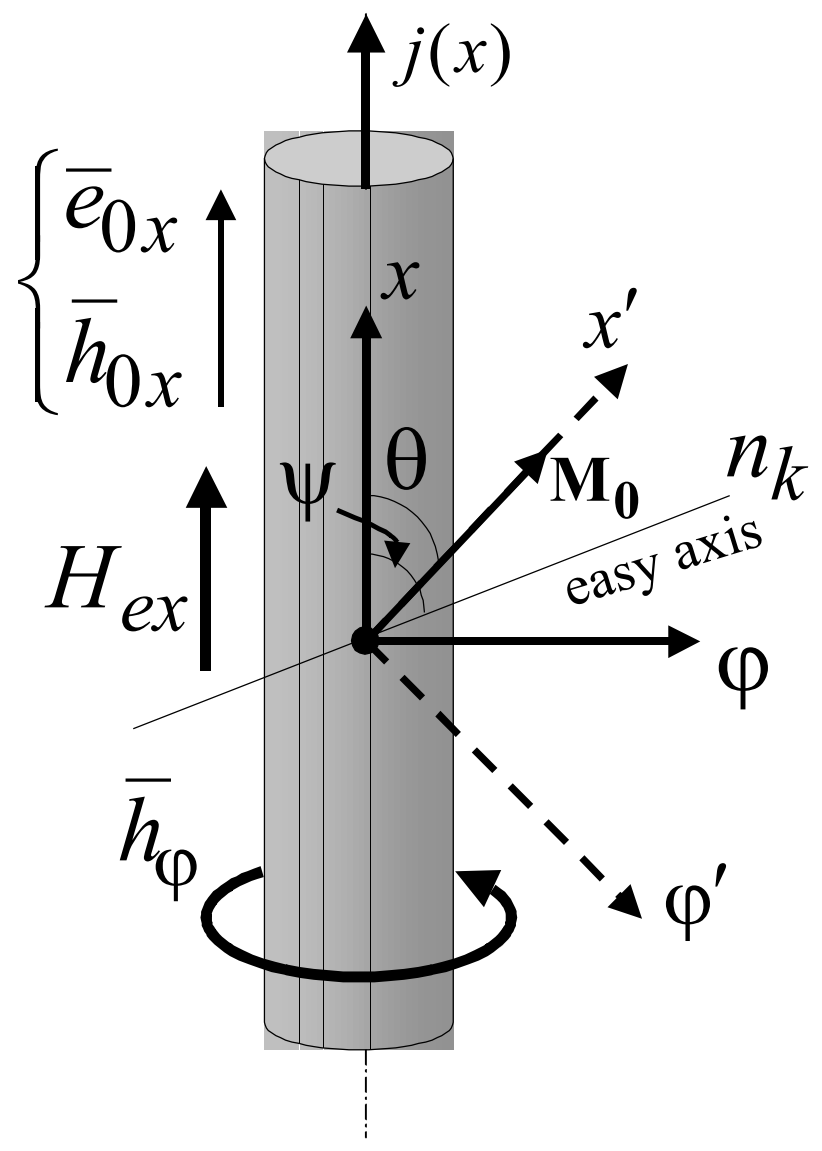

Fig. 3. Schematic diagram of the magnetic configuration in a wire.

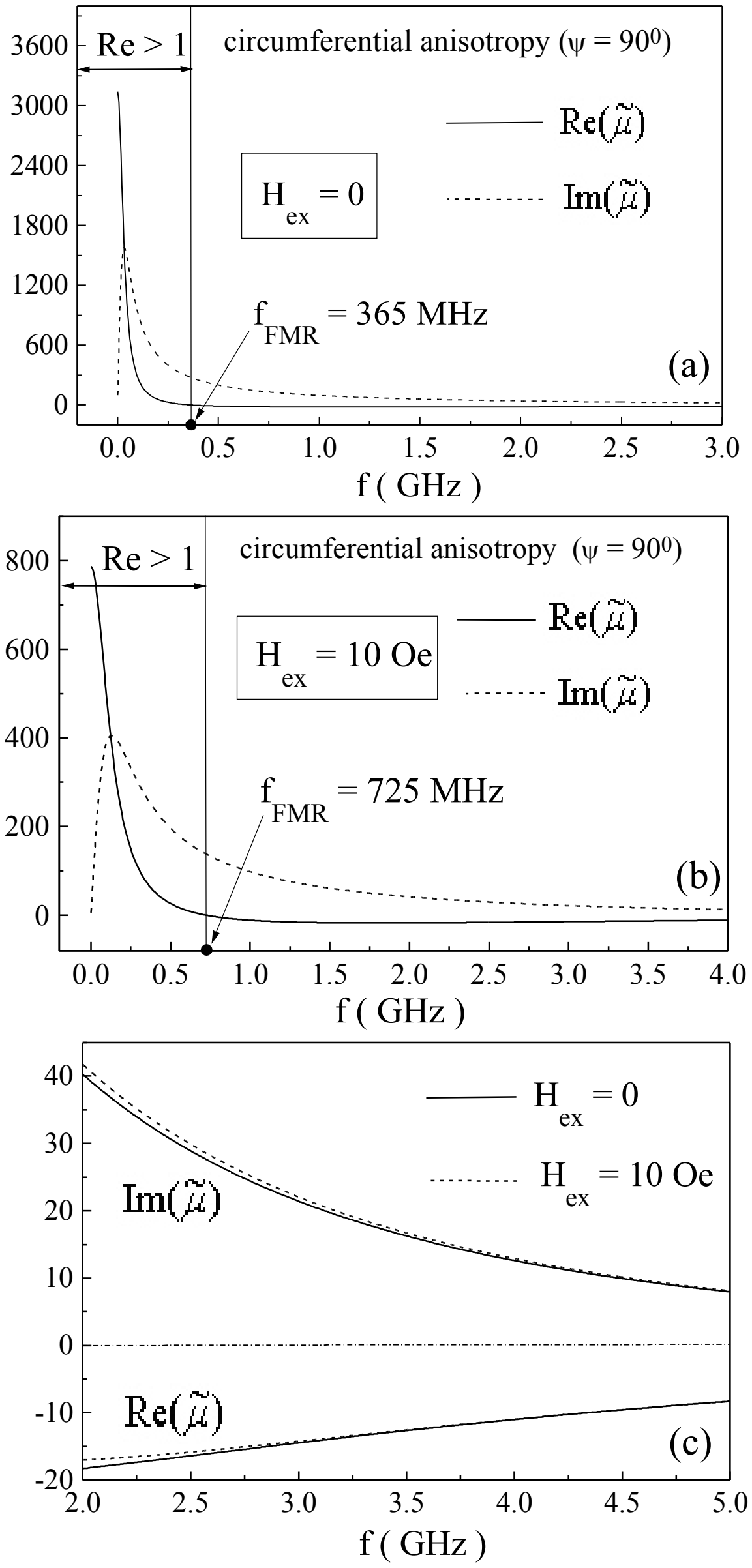

Fig. 4. Typical dispersion curves of the effective permeability $\tilde{\mu}$ for the different external dc magnetic fields $H_{ex}$. Magnetic parameters: anisotropy field $H_K = 2$ Oe, saturation magnetization $M_0 = 500$ G, and gyromagnetic constant $\gamma = 2\times10^7$ (rad/s)/Oe. For frequencies much higher than $f_{FMR}$ (GHz range) $\tilde{\mu}$ becomes insensitive to $H_{ex}$ as shown in (c).

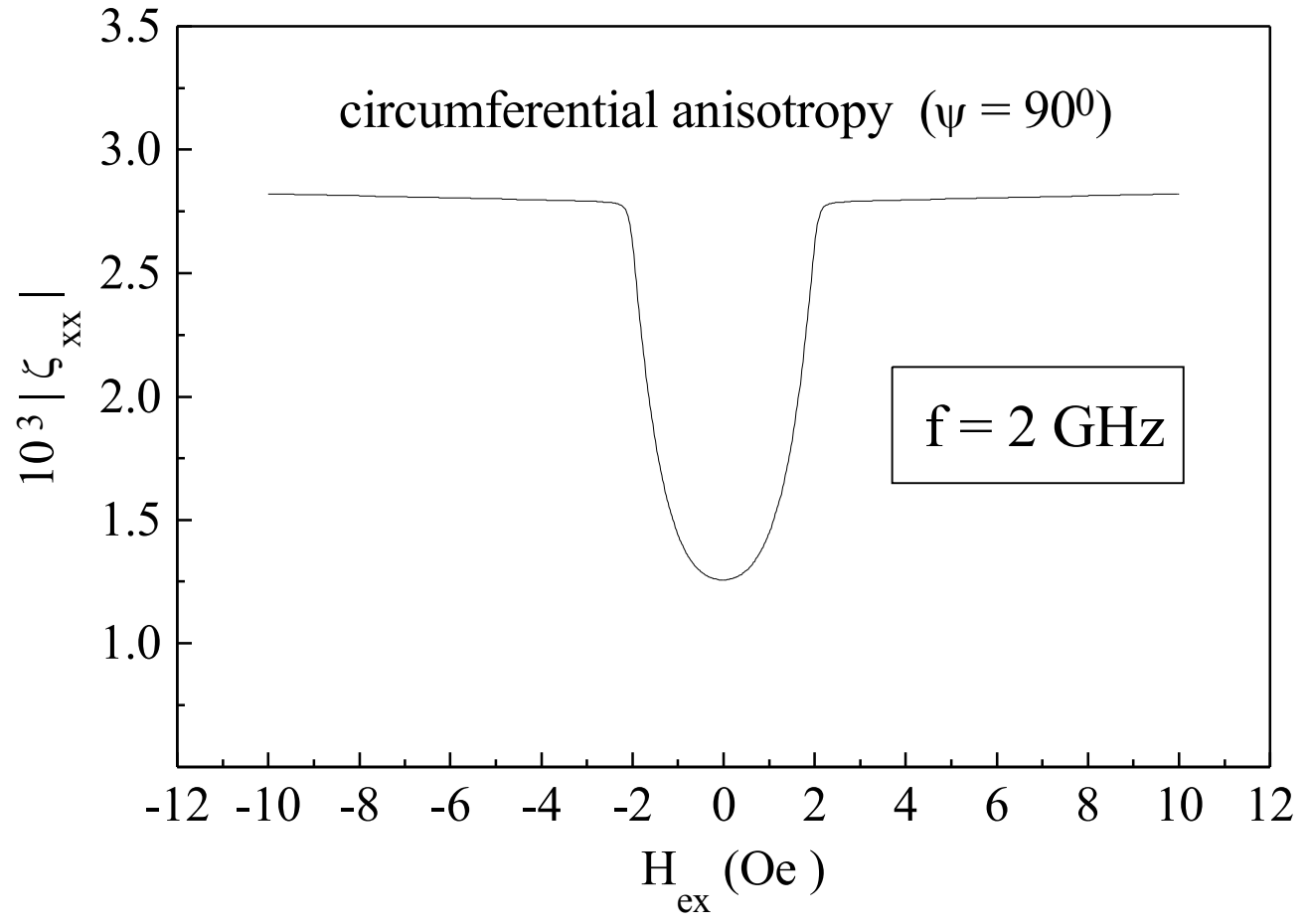

Fig. 5. Typical "valve-like" field dependence of the longitudinal impedance $|\varsigma_{xx}(H_{ex})|$ in the GHz range: $|\varsigma_{xx}(H_{ex})|$ is approximately constant for $|H_{ex}| \geq H_K$ reflecting the field dependence of $\cos^2(\theta)$ since $\tilde{\mu}$ looses its field sensitivity (as shown in Fig. 4(c)).

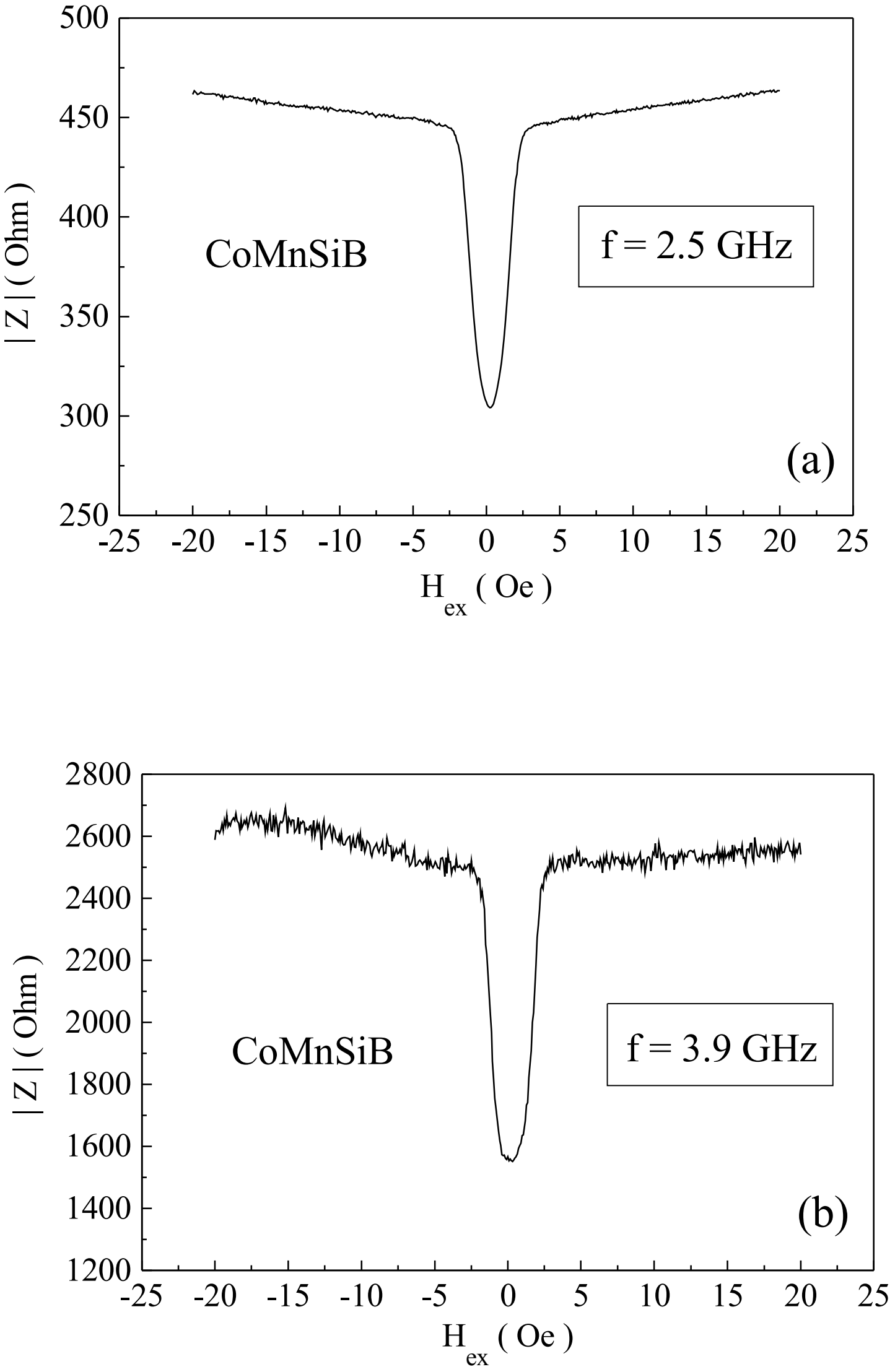

Fig. 6. "Valve-like" field dependences of the impedance $|Z(H_{ex})|$ measured at GHz frequencies in a CoMnSiB amorphous glass-coated wire of $\sim 11-12$ μm in diameter with vanishing negative magnetostriction $\sim -2\times10^{-7}$, the anisotropy field $H_K \sim 2.5$ Oe and the coercitivity field $H_c \sim 0.1-0.15$ Oe.

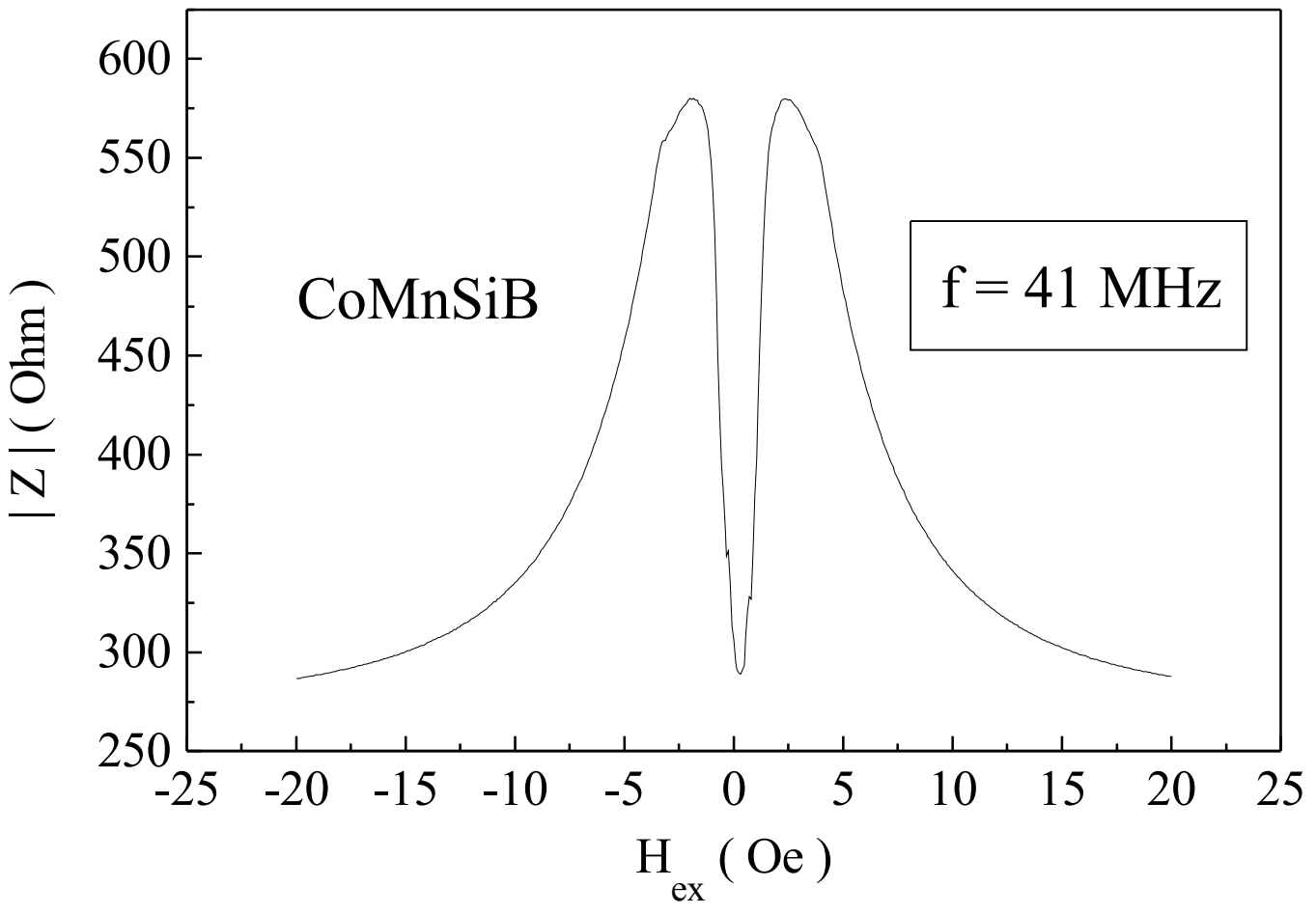

Fig. 7. Typical field dependence of the impedance $|Z(H_{ex})|$ at MHz frequencies in a Co-based wire with a circumferential anisotropy. The field dependence was measured for the same CoMnSiB amorphous glass-coated wire. The impedance $|Z(H_{ex})|$ has two well-defined peaks at $H_{ex} \approx \pm H_K$ and decreases with $|H_{ex}| > H_K$ reaching the level of $\sim|Z(0)|$. This field behavior completely differs from that in the GHz range, as shown in Figs. 5 and 6.

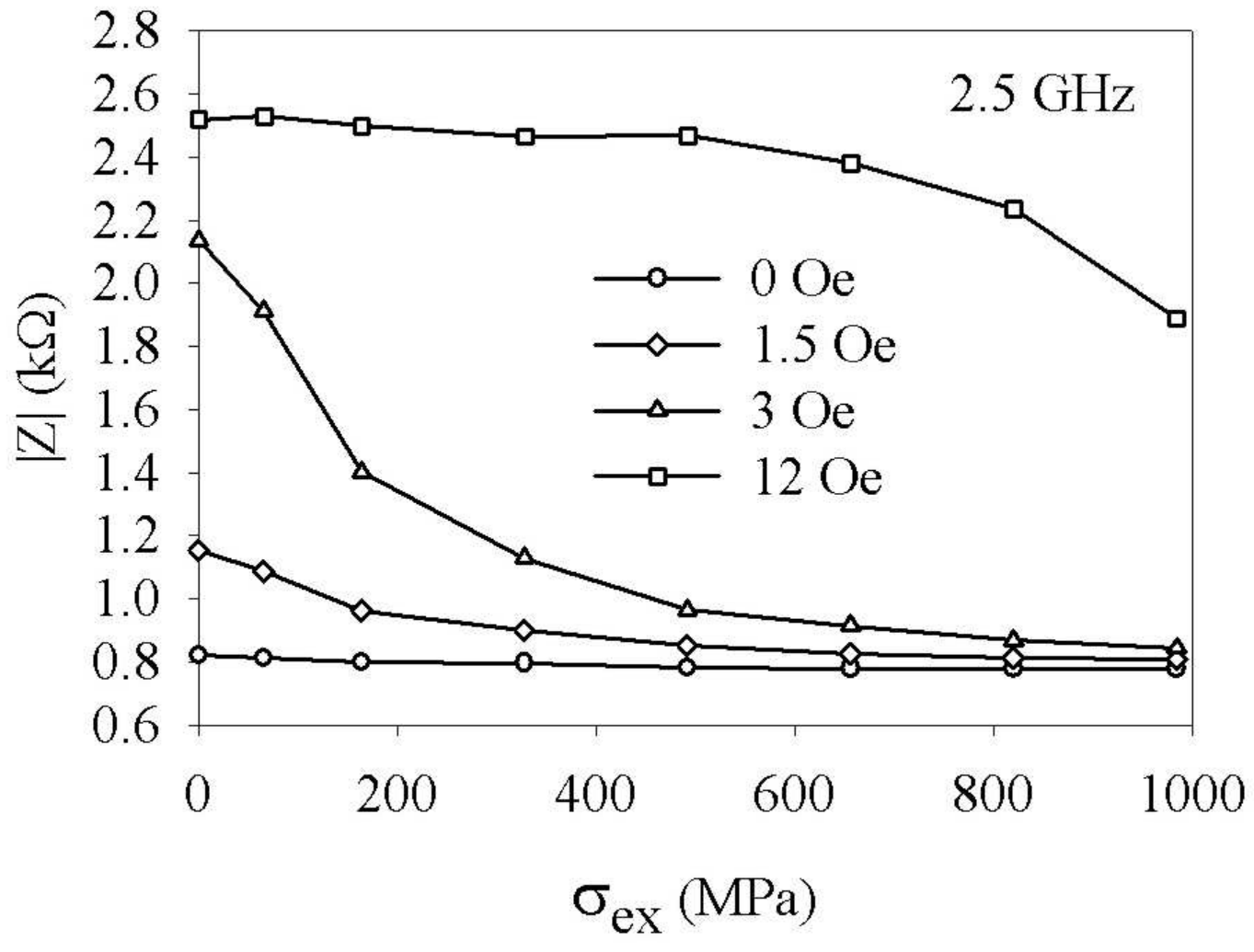

Fig. 8. Experimental plots of the impedance $|Z(\sigma_{ex})|$ vs. applied tensile stress $\sigma_{ex}$ with the external dc magnetic field $H_{ex}$ as a parameter. The highest stress sensitivity is obtained for $H_{ex} \sim H_K$, which is sufficient to saturate the wire.

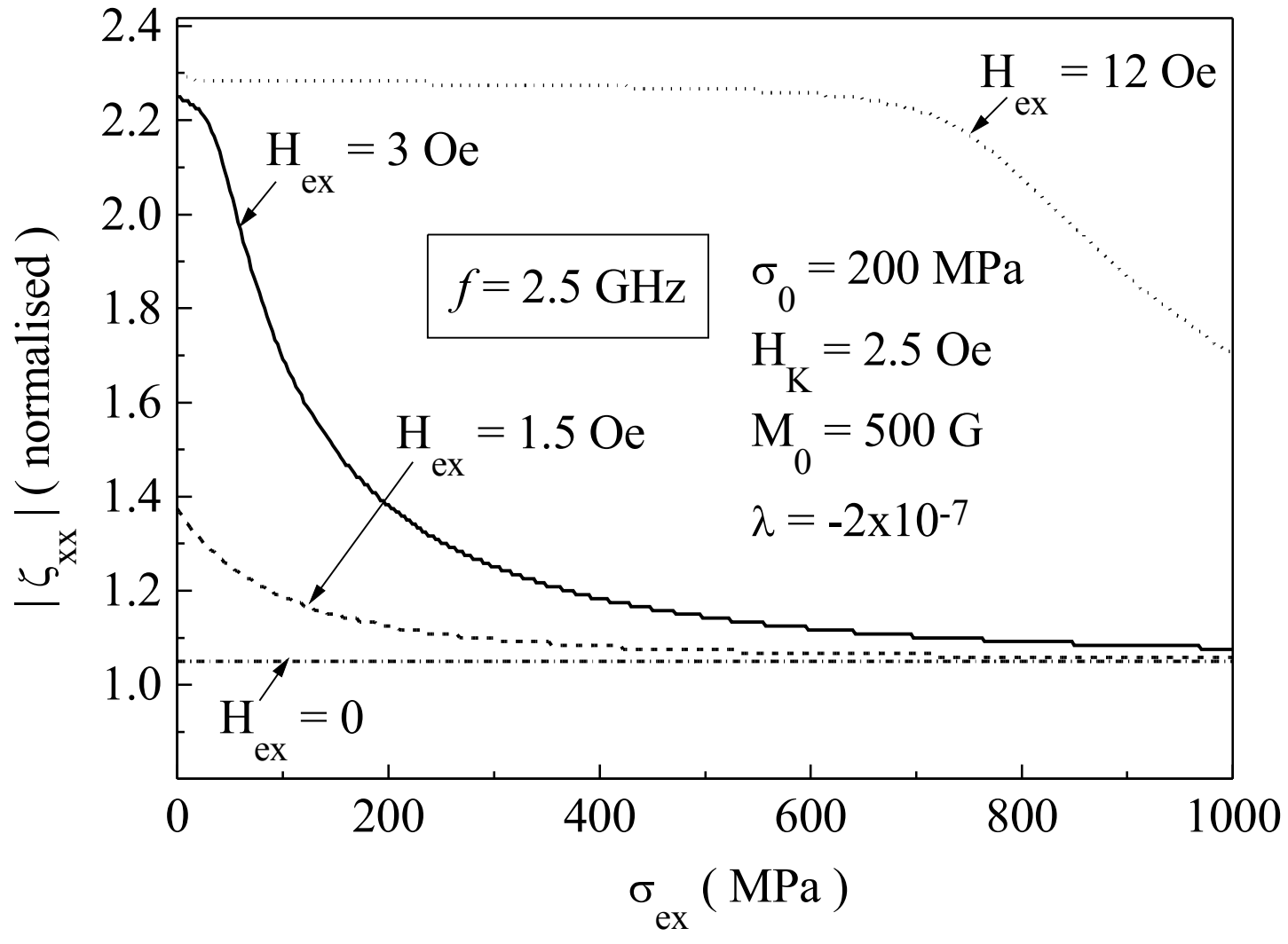

Fig. 9. Theoretical plots of the impedance $|Z(\sigma_{ex})|$ vs. applied tensile stress $\sigma_{ex}$ with the external dc magnetic field $H_{ex}$ as a parameter. The parameters used for calculation are $\lambda = -2 \cdot 10^{-7}$, $M_0 = 500$ G, $\sigma_0 = 200$ MPa. The theoretical curves demonstrate very good agreement with the experimental results in Fig. 8.

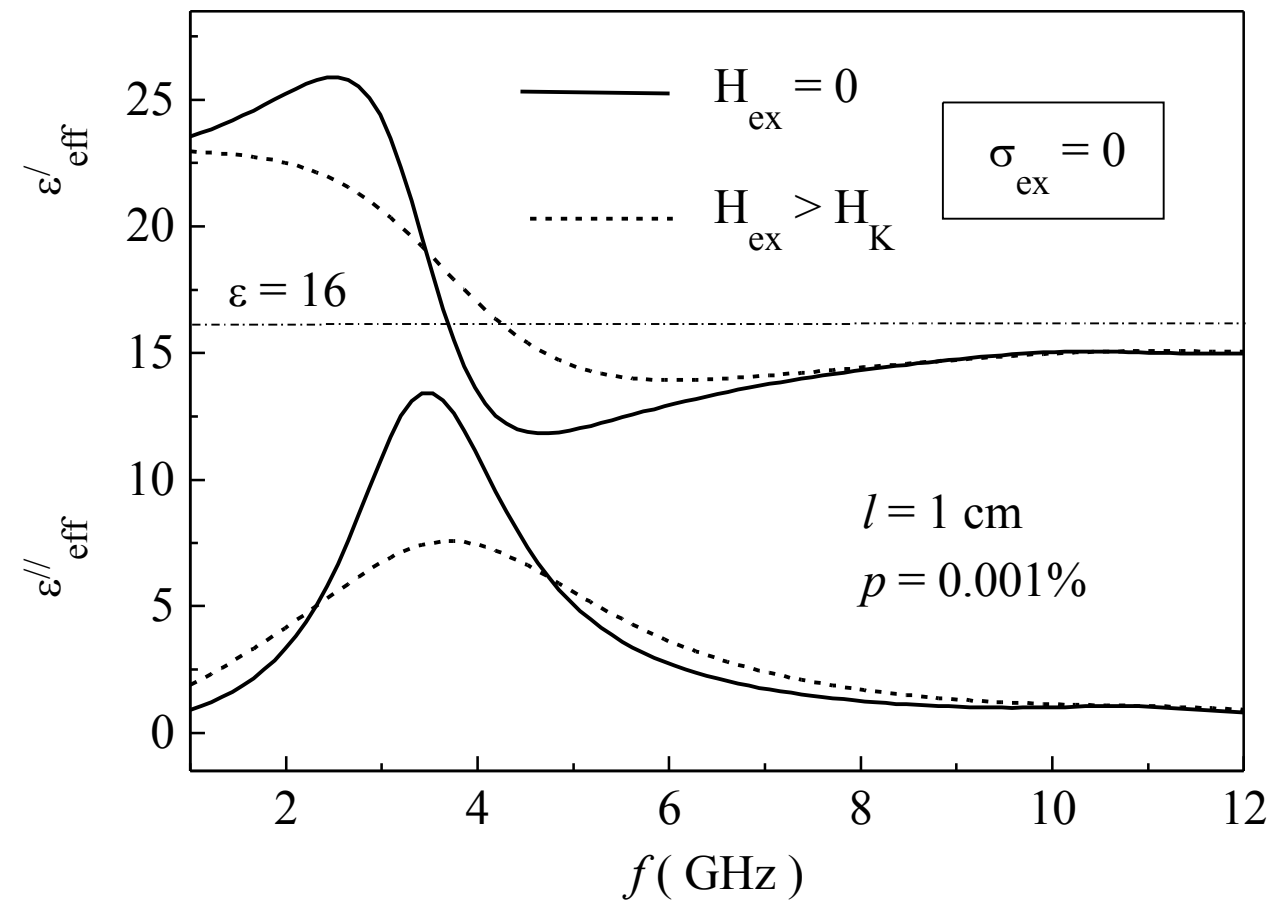

Fig. 10. Transformation of the dispersion of the effective permittivity $\varepsilon_{eff}(\omega)$ from a resonance type to a relaxation one due to $H_{ex}$ calculated in the vicinity of the antenna resonance. The external tensile stress $\sigma_{ex}$ is zero. The composite has the microstructure shown in Fig. 2. The inclusion volume concentration is $p = 0.001\%$ and matrix permittivity is $\varepsilon = 16$.

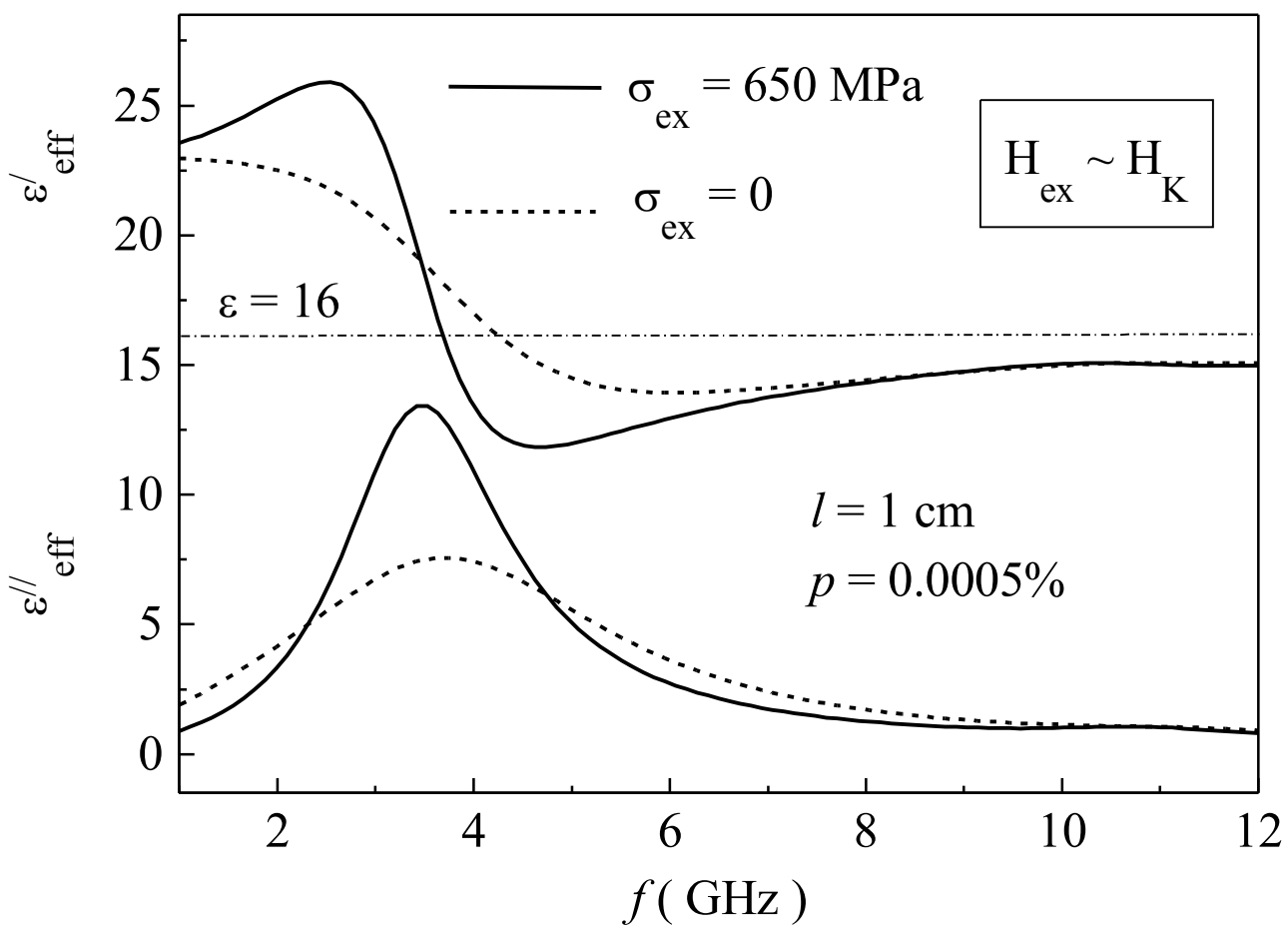

Fig. 11. Transformation of the dispersion of the effective permittivity $\varepsilon_{eff}(\omega)$ from a resonance type to a relaxation one due to the external tensile stress $\sigma_{ex}$ calculated in the vicinity of the antenna resonance. The composite sample is composed of the ordered wire inclusions embedded into a dielectric matrix. The external tensile stress $\sigma_{ex}$ is transmitted to each wire inclusion through the composite matrix. The external dc magnetic field $H_{ex}$ is applied along the wire inclusions and it has the fixed value $H_{ex} \sim H_K$ providing the maximum stress sensitivity (see Figs. 8 and 9). The inclusion volume concentration is $p = 0.0005\%$ and matrix permittivity is $\varepsilon = 16$.

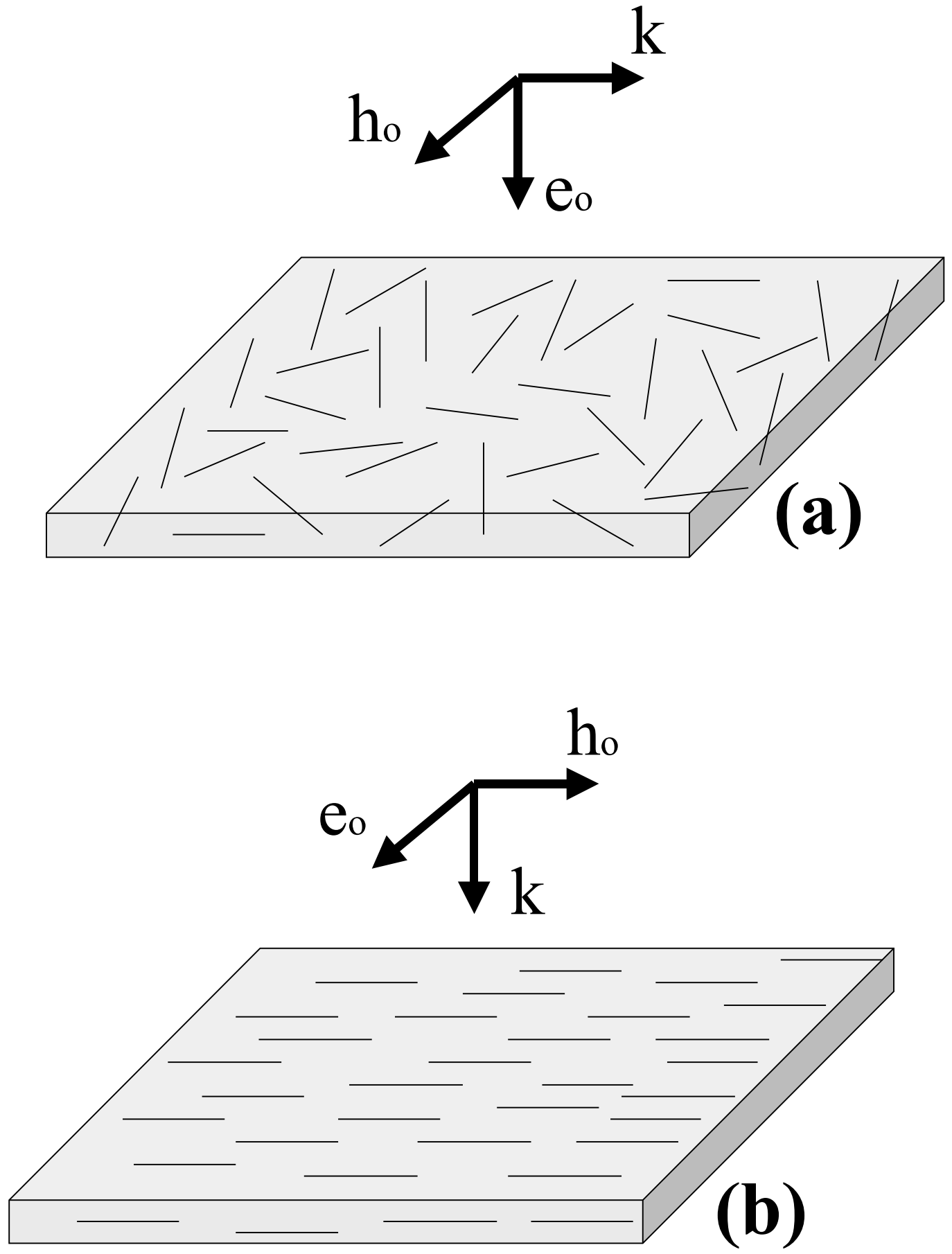

Fig. 12. Composite microstructures and polarizations of the electromagnetic wave providing the magnetic excitation in the composite layer with the random orientations of the wire inclusions in (a) and the ordered wire inclusions in (b).